\documentclass[letterpaper,12pt]{article}

\usepackage{amsfonts,amsmath,graphicx,multirow,rotating,dsfont,color,natbib,enumerate}
\usepackage[latin1]{inputenc}
\usepackage{setspace}

\newtheorem{thm}{Theorem}[section]
\newtheorem{lem}{Lemma}[section]
\newtheorem{prop}{Proposition}[section]

\newtheorem{rem}{Remark}[section]

\def\1{{{\mbox{$\mathds{1}$}}}}

\newcommand{\E}{\mathbb{E}}
\newcommand{\R}{\mathbb{R}}

\def\N{\mathbb{N}}

\def\rmd{\mathrm{d}}

\begin{document}

\begin{center}
{\Large \textbf{Adaptive Density Estimation in the Pile-up Model Involving  Measurement Errors}}\\
~\\~\\~\\
Fabienne Comte, Tabea Rebafka\footnote{Fabienne Comte is statistician at MAP5, UMR 8145 CNRS, Paris Descartes University, France, email: fabienne.comte@parisdescartes.fr. Tabea Rebafka is statistician at LPMA, University of Paris 6, UPMC, France, email: tabea.rebafka@upmc.fr. The authors wish to thank PicoQuant GmbH, Berlin, Germany for kindly providing the TCSPC data. }
\end{center}

\begin{abstract}
Motivated by fluorescence lifetime measurements this paper considers the problem of
nonparametric density estimation in the pile-up model.
Adaptive nonparametric estimators are proposed for the pile-up model in its simple form as well as in the case of
additional measurement errors. Furthermore,  oracle type risk bounds for the
mean integrated squared error (MISE) are provided. Finally, the estimation methods are assessed by a simulation study and the application to real fluorescence lifetime data.
\end{abstract}


%\doublespacing

\section{Introduction}
This paper is concerned with nonparametric density estimation in a specific inverse problem. Observations are not directly available from the target distribution, but suffer from both measurement errors and the so-called pile-up effect. The pile-up effect refers to some right-censoring, since an observation is defined as the minimum of a random number of i.i.d. variables from the target distribution. The pile-up distribution is thus the result of a nonlinear distortion of the target distribution.
In our setting we also take into account  measurement errors, that is the pile-up effect applies to the convolution of the target density and a known error distribution. The aim is to estimate the target density in spite of the pile-up effect and additive noise.

The pile-up model is encountered in time-resolved fluorescence when lifetime measurements are obtained by the technique called
Time-Correlated Single-Photon Counting (TCSPC) \citep{OCP}. The fluorescence lifetime is the
duration that a molecule stays in the excited state before emitting a fluorescence photon \citep{LAKO, VAL}. The distribution of the fluorescence lifetimes associated with a sample of molecules provides precious information on the underlying molecular processes. Lifetimes are used in various applications as e.g. to determine the speed of rotating molecules or to measure molecular distances. This means that the knowledge of the lifetime distribution is required to obtain information on physical and chemical processes.

In the TCSPC technique, a short laser pulse excites a random number of molecules, but for technical reasons, only the arrival time of the very first fluorescence photon striking the detector can be measured, while the arrival times of the other photons are unobservable. The arrival time of a  photon  is the sum of the fluorescence lifetime and some noise, which is some random time due to the measuring instrument as e.g. the  time of flight of the photon in the photon-multiplier tube. Hence, TCSPC observations can be described by a pile-up model with measurement errors. The goal is to recover the distribution of the lifetimes of \textit{all} fluorescence photons from the piled-up observations.

Until recently TCSPC was operated in a mode where the pile-up effect is negligible. However, a shortcoming of this mode is that the acquisition time is very long.  Recent studies  have made clear that from an information viewpoint it is a better strategy to operate TCSPC in a mode with considerable pile-up effect \citep{RRS1,RRS2}. Consequently, an estimation procedure is required that takes the pile-up effect into account. The concern of this paper is to provide such a nonparametric estimator of the target density and furthermore to include  measurement errors in the model in order to deal with  real fluorescence data.
 Therefore, we develop adequate deconvolution strategies for the correction in the pile-up model and test those methods on simulated data as well as on real fluorescence data.

It is noteworthy that the pile-up model is connected to survival analysis, since it can be considered as a special case of the nonlinear transformation model \citep{tsodikov03}.  Indeed, it is straightforward to extend the methods proposed in this paper to the more general case.
Moreover, the model can also be viewed as a biased data problem with known bias \citep{BCG}. As a consequence, the first part of the study is rather classical.
Nonetheless, the consideration of measurement errors  in the second part is new and fruitful.
Indeed, we show that deconvolution methods can be used to complete the study in the spirit of \cite{CRT}. These techniques are of unusual use in both  survival analysis and pile-up model studies. Numerical results confirm the adequacy of these methods in practice.

In Section \ref{firststep} a nonparametric estimation strategy for the pile-up model (without measurement errors) is presented to recover the target density.
More precisely, a projection estimator is developed based on finite dimensional functional spaces and a tool is proposed to automatically select the model dimension achieving the best possible rate of convergence.
In Section \ref{secondstep}  additional measurement errors are taken into consideration leading to an estimator based on  Fourier deconvolution methods. The rates obtained in this framework depend on the smoothness of the error density and on the choice of a cut-off parameter. Furthermore, a cut-off selection strategy is proposed to achieve an adequate bias-variance trade-off. In Section~\ref{Simu} the performance of the methods is assessed via
 simulations and by an application  on a dataset of fluorescence lifetime measurements. All proofs are relegated to Section \ref{proofs}.

\section{Nonparametric Estimator for the Pile-up Model}\label{firststep}
This section introduces the pile-up model and presents the nonparametric estimation approach in the easier setting of the pile-up model before extending it in Section~\ref{secondstep} to the pile-up model including additive noise.

\subsection{The pile-up model}
Let $\{Y_k, k\geq 1\}$ be a sequence of independent positive random
variables with target probability density function (pdf) $f_Y$ and cumulative distribution function (cdf) $F$. Moreover, let $N$ be a random variable taking its values in ${\mathbb N^*} = \{1,2,\dots\}$ independently of this
sequence. Then an observation of the  pile-up  model is distributed as the random variable $Z$ taking values in ${\mathbb R}_+$ defined by
$Z= \min\{Y_1,\dots, Y_N\}.$
In \cite{RRS1} it is shown that the cdf$G$ of $Z$, referred to as the \textit{pile-up distribution function}, is given by
\begin{equation}\label{repftog}
G(z)=1-M(1-F(z)), \;\; z\in {\mathbb R}_+\;,
\end{equation}
where $M$ is the probability generating function associated with $N$ defined as
$M(u)={\mathbb E}(u^N)$ for $u\in [0,1].$
Moreover, if $F$ admits a density $f_Y$ with respect to the  Lebesgue measure on ${\mathbb R}_+$, then $G$ admits a density $g$. Denoting $\dot{M}(u)={\mathbb E}(Nu^{N-1})$, $\ddot{M}(u)= {\mathbb E}(N(N-1)u^{N-2})$ for all $u\in [0,1]$, the pile-up density $g$ is given by
\begin{equation}\label{ftog}
g(z)= f_Y(z) \dot{M}(1-F(z))\;,\quad  z\in {\mathbb R}_+\;.
\end{equation}
Note that the generating function $M:[0,1]\to[0,1]$ is bijective for any distribution of $N$ and we denote its inverse function by $M^{-1}$.
If $\E[N^2]<\infty$ and ${\mathbb P}(N=1)\neq 0, {\mathbb P}(N=2)\neq 0$, then the functions $\dot M$ and $\ddot M$ are bounded by some constants $0<a<b<+\infty$ satisfying
\begin{equation}\label{cw}
a<\dot{M}(u)< b \quad\mbox{  and }  \quad a<\ddot{M}(u) <b \qquad \text{for all } u\in [0,1]\;.
\end{equation}

\begin{rem} {\rm In the more general nonlinear transformation model the function $M:[0,1]\to[0,1]$ in (\ref{repftog}) is not necessarily a probability generating function, but any function $M$ such that $G$ given by (\ref{repftog})  is a cdf \citep{tsodikov03}. That is  $G$ is still the result of a distortion of the target distribution $F$, but the interpretation as a  minimum is no longer valid. Those models are studied in survival analysis. The estimators proposed in this paper for the pile-up model are also applicable for nonlinear transformation models.}
\end{rem}

\noindent {\bf Main example.}
In the fluorescence application it is   assumed that the number $N$ of photons per excitation cycle follows a Poisson distribution with known parameter $\mu$. Note that the events where no photon is detected, i.e. $N=0$, are discarded. Hence, we consider a Poisson distribution restricted on $\N^*$ with renormalized probability masses given by
${\mathbb P}(N=k)={\mu^k}{/k!/ (e^{\mu}-1)}.$
As $\mu$ is supposed to be known, the functions $M$ and $\dot{M}$ are known as well and given by
$M(u)= (e^{\mu u}-1) /(e^{\mu}-1)$  and $\dot{M}(u)=\mu e^{\mu u}/(e^{\mu}-1)$.

\subsection{Estimator of the target density in the pile-up model}\label{subsec_estim_pileup}
The goal is to estimate the target density $f_Y$ from i.i.d.observations $Z_1,\dots,Z_n$ of the pile-up distribution $G$. We propose a nonparametric estimator by searching in a collection of functions the one that best fits the data or, in other words, the orthogonal projection of $f_Y$ onto the function space. If $S$ is an adequate subspace of $L^2$, the orthogonal projection of $f_Y$ on $S$ in the $L^2$-sense is the minimizer of $\|f_Y-h\|^2$  for $h$ in $S$, or equivalently, the minimizer of $\|h\|^2-2\langle h, f_Y\rangle$.

As $\langle h, f_Y\rangle= {\mathbb E}(h(Y))$, we need an approximation of moments $\E[h(Y)]$ based on pile-up observations.
We note that inverting relation (\ref{repftog}) gives  $1-F(z)=M^{-1}(1-G(z))$. Plugging  this relation into~(\ref{ftog}), we obtain
\begin{equation*}\label{densite}
f_Y(z)=\frac{g(z)}{\dot{M}(M^{-1}(1-G(z)))}= w\circ G(z)~g(z)\quad \mbox{ with } \quad w(u)=\frac 1{\dot{M}(M^{-1}(1-u))}\;.
\end{equation*}

This allows us to relate moments of the target distribution $F$ with moments of the pile-up distribution $G$. More precisely, for any bounded function $h$ the following equality holds
\begin{equation}\label{pileup_property}
{\mathbb E}[h(Y)]={\mathbb E}\left[h(Z)~w\circ G(Z)\right]\;.
\end{equation}
To construct an estimator of the moment $\E[h(Y)]$ based on  pile-up observations, relation (\ref{pileup_property}) suggests to replace the distribution function $G$ by its empirical version $\hat G_n(z)= \sum_{i=1}^n \1_{\{Z_i\leq z\}}/n.$
Then an estimator of $\E[h(Y)]$ is given by
\begin{equation}\label{empirical_weighted_mean}
\frac 1n\sum_{i=1}^n h(Z_i)~w\circ\hat G_n(Z_i) = \frac 1n\sum_{i=1}^n h(Z_{(i)})w(i/n)\;,
\end{equation}
as $w\circ\hat G_n(Z_{(i)}) = w(i/n)$ and where $Z_{(i)}$ denotes the $i$-th order statistic associated with $(Z_1,\dots,Z_n)$ satisfying $Z_{(1)}\leq\dots\leq Z_{(n)}$. In the literature such weighted sums of order statistics are known as $L$-statistics.

The approximation of moments $\E[h(Y)]$ by an $L$-statistic is the key property used in the nonparametric estimation strategy that is proposed in the following. In the pile-up model the weights $w(i/n)$ can be viewed as ``corrections" of the observations $Z_i$ as they do not follow the target distribution $F$, but the pile-up distribution $G$.
The weights are bounded because inequality (\ref{cw}) ensures that there exist  constants $w_0, w_1$  such that
\begin{equation}\label{hypw}
\forall u\in [0,1], \;\; 0<w_0 \leq w(u) \leq w_1 <\infty\;.
\end{equation}
The computation of the estimator in (\ref{empirical_weighted_mean}) requires the knowledge of the weight function $w$, which is entirely determined by the distribution of $N$. Hence, in the example above on the Poisson distribution $w$ writes
\begin{equation}\label{wpoisson}
w(u) = \frac{1-e^{-\mu}}{\mu(u(e^{-\mu}-1) +1)}\;,
\end{equation}
with corresponding constants $w_0= (1-e^{-\mu})/\mu$ and $w_1 = (e^\mu-1)/\mu$.

A standard estimation approach of the target density $f_Y$  consists in approximating the orthogonal projection of $f_Y$ onto some function space. More precisely, we suppose that the restriction of $f_Y$ on some interval $A$ is square integrable, i.e. $f_Y\1_A\in\mathbb{L}^2(A)$.
For a given orthonormal sequence  $(\varphi_\lambda)_{\lambda \in\Lambda_m}$ in ${\mathbb L}^2(A)$ define the subspace $S_m= {\rm Span}(\varphi_\lambda, \lambda \in \Lambda_m).$
The cardinality of $\Lambda_m$ (which is also the dimension of $S_m$) is denoted by $D_m$ and supposed to be finite.

By using the moment estimator proposed in (\ref{empirical_weighted_mean}), an approximation of the projection of $f_Y$ onto $S_m$ can be defined as
$$\hat f_m=\arg\min_{h\in S_m}\gamma_n(h) \qquad \text{ with } \quad
\gamma_n(h)=\|h\|^2-\frac 2n\sum_{i=1}^n h(Z_{(i)})~w(i/n)\;,$$
since $\gamma_n(h)$ is an estimator of $\|h\|^2-2\E[h(Y)]$. Note that the explicit formula of the estimate is given by
\begin{equation}\label{estimbias}
\hat f_m=\sum_{\lambda\in \Lambda_m} \hat a_\lambda \varphi_\lambda \qquad \mbox{ with } \quad\hat a_\lambda=\frac 1n \sum_{i=1}^n \varphi_\lambda(Z_{(i)}) w(i/n)\;.\end{equation}

For this estimator  the following risk bound is shown in Section \ref{proofs}.
\begin{prop}\label{risksansbruit}
Let $f_m$ be the orthogonal projection in the ${\mathbb L}^2$-sense of $f_Y$ on $S_m$. Assume that (\ref{hypw}) holds and that $w$ is Lipschitz continuous, i.e.
\begin{equation}\label{lip}
\mbox{there exists }  c_w>0 \mbox{ such that }  |w(x)-w(y)|\leq c_w |x-y|\;.
\end{equation}
 Assume moreover that
\begin{equation}\label{phi0}
\mbox{there exists } \; \Phi_0>0 \;  \text{ such that } \; \|\sum_{\lambda\in\Lambda_m}\varphi_{\lambda}^2\|_{\infty}\leq \Phi_0 D_m\;,
\end{equation}
then
\begin{equation}\label{risk1}
{\mathbb E}(\|\hat f_m-f_Y\1_A\|^2) \leq \|f_Y\1_A-f_m\|^2 + C\frac{D_m}n\;,\end{equation}
where $C$ depends on $\Phi_0$, $w_1$ and the Lipschitz constant $c_w$ of $w$.
\end{prop}

\noindent \begin{rem} {\rm
It follows from equation (\ref{cw})  that the Lipschitz constant $c_w$  verifies  $c_w\leq b/a^3$  since $w'(u)=\ddot{M}\circ
M^{-1}(1-u)/[\dot{M}\circ M^{-1}(1-u)]^3$. In the  Poisson example where $w$ is given by (\ref{wpoisson})
we have $c_w=(e^\mu-1)^2/\mu$.
}\end{rem}

\subsection{Examples of model collections }
Our goal is the estimation of $f_Y$ in a nonparametric setting without knowledge of the best approximation space. Instead of a single space $S_m$, we rather consider a collection $\{S_m,m\in {\mathcal M}_n\}$ of models  and we thus have to face the problem of model selection. Before presenting an estimator of the model $m$, we give some illustrating examples of model collections  $S_m$ and we discuss some general conditions for the approximation spaces under which our estimation approach performs well.

In the following $A$ is supposed to be a compact set. For simplicity, we set $A=[0,1]$. \\
\noindent $[{\rm T}]$ {\em Trigonometric spaces} $S_m$ are generated
by the functions
$$\{1, 2^{1/2}\cos(2\pi jx), 2^{1/2}\sin(2\pi j x) \text{ for }j=1,\dots, m \}\;.$$
The dimension of $S_m$ is $D_m=2m+1$  and we may take $m\in {\mathcal M}_n=\{1, \dots, [n/2]-1\}$.\\
\noindent $[{\rm DP}]$ {\em Dyadic piecewise polynomials spaces} of degree $r$ on the partition of $[0,1]$ given by the subintervals $I_j=[(j-1)/2^p, j/2^p]$ for $j=1, \dots, 2^p$, see \cite{BM}, Section 4.2.2.

\noindent $[{\rm W}$] {\em Dyadic wavelet generated spaces} with regularity
$r$ and compact support, see e.g. \cite{DAU, DJKP}.

We now give the key properties that a general model collection $\{S_m,m\in {\mathcal M}_n\}$  must  fulfill to fit into our framework.
\begin{enumerate}
\item[(${\mathcal{H}}_1$)] Norm connection:
$\{S_m, m\in {\mathcal M}_n\}$ is a collection of
finite dimensional linear sub-spaces of ${\mathbb L}^2([0,1])$
with dimension dim$(S_m)=D_m$ satisfying $D_m\leq N_n\leq  n$, $\forall m\in
{\mathcal M}_n$ and
\begin{equation}
\label{connectionnorme}
\mbox{ There exists }  \Phi_0>0 \mbox{ such that } \|t\|_{\infty} \leq \Phi_0 D_m^{1/2}\|t\|, \mbox{ for all } m\in{\mathcal M}_n,  t\in S_m.
 \end{equation}
\end{enumerate}
Let $(\varphi_{\lambda})_{\lambda \in \Lambda_m}$ be an orthonormal basis of $S_m$, where $|\Lambda_m|=D_m$.
It follows from Birg\'e and Massart~(1997) that Property (\ref{connectionnorme})
in the context of (${\mathcal{H}}_1$) is equivalent to (\ref{phi0}) for all $m\in {\mathcal M}_n$.
This condition is easily checked for collection [T] with $\Phi_0=1$.
For collection [DP] see a detailed description in \cite{BM}, Section 2.2, showing that condition (\ref{phi0})  holds  with $\Phi_0^2=r+1$.
It is known that (\ref{phi0}) is also satisfied for wavelet bases [W].

Additionally, for results concerning adaptive estimators the following assumption is required.
\begin{enumerate}
\item[(${\mathcal{H}}_2$)] Nesting condition: $\{S_m, m\in {\mathcal M}_n\}$ is a collection of models such that there exists a space  ${\mathcal S}_n$  belonging to the collection such that $S_m\subset {\mathcal S}_n$ for all $m\in {\mathcal M}_n$.
Denote by $N_n$ the dimension of ${\mathcal S}_n$, i.e. dim$({\mathcal S}_n)=N_n\leq n$.
\end{enumerate}
This condition ensures that $D_m \leq N_n$ for all $m \in {\mathcal{M}}_n$.\\

Another key property of those spaces lies in the bias evaluation. Indeed, if we assume that $f_Y\1_A=f_A$ belongs to
a ball of some Besov space  ${\mathcal B}_{\alpha,2,\infty}(A)$ with
$r+1\geq\alpha$, then for $\|f_A\|_{\alpha, 2,\infty}\leq L$ we have $\|f_A-f_m\|^2 \leq C(\alpha,L)
D_m^{-2\alpha}$ \citep[][Lemma 12]{BBM}. Thus, choosing $D_{m^*}=O(n^{1/(2\alpha+1)})$ in Inequality (\ref{risk1}) yields that the  mean square risk satisfies ${\mathbb E}(\|\hat
f_{m^*}-f_A\|^2)\leq O(n^{-2\alpha/(2\alpha+1)})$. This rate is known to be optimal  in the minimax sense for density estimation for direct observations \citep{DJKP}.

\subsection{Adaptive estimator}
From the risk bound (\ref{risk1}) it is clear  that a bias-variance trade-off must be achieved. The idea consists in searching the model $m$ that minimizes the risk bound (\ref{risk1}).
As $\|f_Y-f_m\|^2=\|f_Y\|^2 -\|f_m\|^2$, this is equivalent to minimize  $-\|f_m\|^2 + CD_m/n$, where the term $-\|f_m\|^2$ can be estimated
by  $-\|\hat f_m\|^2 = \gamma_n(\hat f_m)$.
Consequently, we propose the following model selection device
\begin{equation}\label{hatm}
\hat m=\arg\min_{m\in {\mathcal M}_n} [\gamma_n(\hat f_m) + {\rm pen}(m)]\;,
\end{equation}
where the penalty term pen$(m)$ is of the same order as
the variance, i.e. $CD_m/n$.
Using this approach the following result can be shown.
\begin{thm}\label{adaptbias}  Consider collections {\rm [DP]} or {\rm [W]} with $N_n\leq O(n)$ or collection {\rm [T]} with $N_n\leq O(\sqrt{n})$ and assume that $f_Y$ is bounded on $A$, i.e. $\|f_Y\|_\infty<\infty$.  Let $\hat m$ be defined by (\ref{hatm}) with
\begin{equation}\label{pen} {\rm pen}(m)= \kappa \left(\int_0^1 w^2(u)\rmd u \right) \frac{D_m}n\;.\end{equation}Then there exists a numerical constant $\kappa$ such that we have
\begin{equation}\label{oracle}  {\mathbb E}(\|f-\hat f_{\hat
m}\|^2)\leq C\inf_{m\in {\mathcal M}_n}\left(\|f-f_m\|^2 + \left(\int_0^1 w^2(u)\rmd u\right)\frac{D_m}n\right) + K\frac{\ln^2(n)}n, \end{equation}
where $C$ is a numerical constant and $K$ depends on $c_w$, $\|f_Y\|_{\infty}$ and the basis.
\end{thm}
Risk bounds of the form (\ref{oracle} ) are often called oracle inequality. Note that the last term $c\ln^2(n)/n$ is clearly negligible with respect to the order of the infimum (in particular, in all Besov cases described above).

In practice, the numerical constant $\kappa$ is calibrated by simulation
experiments based on a few samples.
The selection of $\hat m$ in (\ref{hatm}) is numerically easy, since  the values of $\gamma_n(\hat f_m)$ are given by
$ = -\sum_{\lambda\in \Lambda_m}\hat a_\lambda^2$   with $\hat a_\lambda$ is defined in (\ref{estimbias}).

The proof of the theorem relies on Talagrand's inequality and follows
the line of the proof of Theorem 4.2 in \cite{BC}. Therefore, only a sketch of the proof is provided in Section \ref{proofs}.

\section{Pile-up Model with Measurement Errors}\label{secondstep}

In this section we consider the context where the random variables $Y_i$ are affected by  additional measurement errors. More precisely, the observations have the following form
 $Z = \min\{Y_1+\eta_1,\dots,Y_N+\eta_N\},$
where the measurement errors $\eta_i$ are independent of $Y_i$ and have known density $f_\eta$ with support in $\R^+$.  The pdf $f$ of $X=Y+\eta$ is  the convolution of $f_Y$ and $f_\eta$  denoted by $f=f_Y*f_\eta$. We denote by $u^(t)=\int e^{-itx}u(x)dx$ the Fourier transform of an integrable function $u$.

\subsection{Estimation procedure and risk bound}
In the context of piled-up observations with measurement errors, since obviously $f^*_Y=f^*_X/f^*_\eta$, one may consider the  natural plug-in estimator of $f_Y$ given by
$\hat f_{Y,m}(x)={(2\pi)^{-1}} \int_{-\pi m}^{\pi m}
e^{ixu}{\hat f_{\hat m}^*(u)}/{f_\eta^*(u)} \rmd u,$
provided that the Fourier transform of  $\hat f_{\hat m}$ exists.
However, this approach leads to an accumulation of the estimation errors of the two stages. It is known that especially the application of the inverse Fourier transform is particularly unstable. Hence a better solution may be obtained by a direct approach.

To this end we   note that in this set-up the ``pile-up property'' given by (\ref{pileup_property})  holds for $X=Y+\eta$, that is  ${\mathbb E}(h(X))= {\mathbb E}(h(Z)w\circ G(Z))$.
Hence, a direct estimator of the Fourier transform $f^*_X$ is given by
\begin{equation}\label{estim_FTx}
\widehat{f^*_X}(u) = \frac 1n \sum_{k=1}^n e^{-iZ_{(k)} u}~w(k/n)\;,
\end{equation}
and finally an
estimator of the target density $f_Y$ can be defined as
\begin{equation}\label{estimdirect}
\bar f_m(x)= \frac 1{2\pi} \int_{-\pi m }^{\pi m} e^{iux}\frac{\widehat{f^*_X}(u)}{f_\eta^*(u)}  \rmd u\;.
\end{equation}

For this estimator, the following risk bound can be shown.
\begin{prop}\label{borne} Assume that $w$ satisfies (\ref{hypw}) and (\ref{lip}).
Let $f_{Y,m}$ denote the function verifying $f_{Y,m}^*=f_Y^*\1_{[-\pi m, \pi m]}$. Then
\begin{equation}\label{risk}  {\mathbb E}(\|\bar f_m-f_Y\|^2) \leq \|f_Y-f_{Y,m}\|^2 + C\frac{\Delta_\eta(m)}n
\;\; \mbox{ where } \;\; \Delta_\eta(m)= \frac1{2\pi} \int_{-\pi m}^{\pi m} \frac{\rmd u}{|f_\eta^*(u)|^2}\;,
\end{equation}
and $C$ depends on $\int_0^1 w^2(u)\rmd u$ and on the Lipschitz constant $c_w$ of $w$.\\
\end{prop}
Note that
$\|f_Y-f_{Y,m}\|^2=(2\pi)^{-1}\int_{|u|\geq \pi m} |f_Y^*(u)|^2\rmd u$.

Obviously, the variance depends crucially on the rate of decrease to $0$ of $f_\eta^*$ near infinity. For instance,
if  $f_\eta$ is the standard normal density, the variance is proportional to $\int_{|u|\leq
\pi m}e^{u^2/2}\rmd u/n$, whereas for the Laplace distribution
(i.e. $f_\eta(x)=e^{-|x|}/2$) we have $1/f_\eta^*(u)=1+u^2$ and a variance of order $O(m^4/n)$.

\subsection{Other ways to view the estimator}
The estimator $\bar f_m$ can also be derived in a different way. Recall that in Subsection \ref{subsec_estim_pileup} we defined an estimator by minimizing the contrast $\gamma_n(h)$ which is an approximation of $ \|h\|^2-2\E[h(Y)]$. Writing $\E[h(Y)] = \langle h,f\rangle =(2\pi)^{-1} \langle h^*,f^*_Y\rangle=\frac1{2\pi} \langle h^*,f^*_X/f^*_\eta\rangle$ suggests to consider functions $h$ in $\bar S_m=\{ h,\;  \; {\rm support}(h^*)\subset [-\pi m, \pi m]\}$ and the new contrast
\begin{equation*}\label{gamdec}
\gamma_n^\dagger (h) = \|h\|^2 - \frac 1{\pi }\int h^*(-u)
\frac{ \widehat{f^*_X}(u)}{f_\eta^*(u)} \rmd u \;,\end{equation*}
where $\widehat{f^*_X}$ is given by (\ref{estim_FTx}).
Now we can see that the estimator $\bar f_m$ minimizes the contrast $\gamma_n^\dagger$. Indeed, note that
$\bar f_m^*(u)= \widehat{f^*_X}/f_\eta^*(u)~\1_{[-\pi m, \pi m]}(u)$ and thus $\bar f_m\in \bar S_m$. By Parseval's formula $\langle h,\bar f_m\rangle=(2\pi)^{-1}\langle h^*, \bar f_m^*\rangle$. This yields that $\gamma_n^\dagger(h)=\|h\|^2 -2\langle h, \bar f_m\rangle=\|h-\bar f_m\|^2 -\|\bar f_m\|^2$. Therefore,
$\bar f_m=\arg \min_{h\in \bar S_m} \gamma_n^\dagger(h).$

Another expression of the estimator is obtained by describing more precisely the functional spaces $\bar S_m$ on which the minimization is performed.
To that aim, let us define the sinc function and its translated-dilated versions by
\begin{equation}\label{sinc_fn}
\varphi(x)=\frac{\sin(\pi x)}{\pi x} \quad \mbox{ and }
\quad \varphi_{m,j}(x) = \sqrt{m} \varphi(mx-j)\;,
\end{equation}
where $m$ is an integer that can be taken equal to $2^{\ell}$. It is well known
 that $\{\varphi_{m,j}\}_{j \in \mathbb{Z}}$
is an orthonormal basis of the space of square integrable functions
having Fourier transforms with compact support in $[-\pi
m, \pi m]$ \citep[][p.22]{MEY}. Indeed, as $\varphi^*(u)=\1_{[-\pi, \pi]}(u)$, an elementary computation yields that
$\varphi^*_{m,j}(x)= m^{-1/2}{e^{-ixj/m}} \1_{[-\pi m, \pi
m]}(x).$
Thus, the functions $\varphi_{m,j}$ are such that
$\bar S_m = {\rm Span}\{\varphi_{_{m,j}}, \; j\in \mathbb{Z}\}\; =\{h\in \mathbb{L}_2(\mathbb{R}), \mbox{supp}(h^*) \subset[-m\pi,m\pi ]\}.$
For any function $h\in {\mathbb L}_2({\mathbb R})$, let $\Pi_m(h)$ denote the orthogonal projection of $h$ on $\bar S_m$ given by
$\Pi_m(h)=\sum_{j\in {\mathbb Z}} a_{m,j}(h) \varphi_{m,j}$
with $a_{m,j}(h) = \int_{\mathbb R} \varphi_{m,j}(x)h(x)\rmd x$.
As $a_{m,j}(h)= (2\pi)^{-1} \langle \varphi_{m,j}^*,h^*\rangle$, it follows that
$\Pi_m(h)^*= h^*\1_{[-\pi m, \pi m]},$
and thus $f_{Y,m} = \Pi_m(f_Y).$
Since $\bar f_m$ minimizes $\gamma^\dagger_n$,  this yields that the estimator $\bar f_m$ can be written in the following convenient way
\begin{equation}\label{explicit}
\bar f_m
 = \sum_{j\in {\mathbb Z}} \bar a_{m,j} \varphi_{m,j} \quad \mbox{ with }
\bar a_{m,j}= \frac 1{2\pi } \int \varphi_{m,j}^*(-u)\frac{
\widehat{f_X^*}(u)}{f_\eta^*(u)} \rmd u \;.
\end{equation}
Consequently $\|\bar f_m\|^2=\sum_j |\bar a_{m,j}|^2$.

Finally, one can see that  $\sum_{j\in {\mathbb Z} } \varphi_{m,j}^*(u)\varphi_{m,j}(x) = e^{-ixu}\1_{|x|\leq \pi m}$. This is another way to see that
(\ref{explicit}) and (\ref{estimdirect}) actually define the same estimator.

\begin{rem} {\rm An interesting remark follows from equation (\ref{explicit}).
In the case where no noise has to be taken into account, i.e.
$f_\eta^*(u)\equiv 1$, the integral in (\ref{explicit}) becomes $ \int \varphi_{m,j}^*(-u)e^{-iuZ_k} \rmd u=2\pi \varphi_{m,j}(Z_k)$. Hence, $\bar a_{m,j}= (1/n) \sum_{k=1}^n
\varphi_{m,j}(Z_{(k)})w(k/n)$. We recognize the coefficients of the estimators given by formula (\ref{estimbias}) of the setting in Subsection \ref{subsec_estim_pileup}, when
the orthonormal basis $(\varphi_\lambda)_\lambda$ is the sinc basis.}
\end{rem}

\subsection{Discussion on the type of noise}\label{OS}
\begin{figure}[h!]
\begin{center}
\includegraphics[scale=.34]{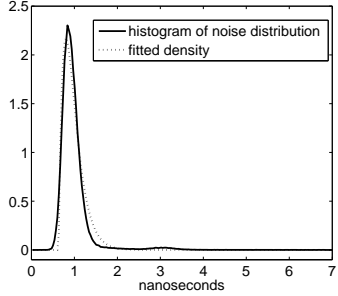} \caption{Normalized histogram based on a sample of the noise distribution (solid line) and the fitted density (dashed line) having the form of (\ref{dens}) with $\hat\alpha=0.961$, $\hat\beta=0.941$, $\hat\nu=5.74$, $\hat\tau=5.89$.}\label{fitted_density}\end{center}
\end{figure}

To determine the rate of convergence of the MISE, it is necessary to specify the type of the noise distribution. Here two cases are considered.
First, the noise distribution can be  exponential with density given by
$ f_\eta(x)=\theta e^{-\theta x}\1_{x>0}\;,$
for some $\theta>0$.
Then we have $f_\eta^*(u)=\theta/(\theta+iu)$, $|f_\eta^*(u)|^2=1/(1+u^2/\theta^2)$ and $\Delta_\eta(m) = m+\pi^2m^3/(3\theta^2)$.

In the fluorescence setting,  we found that TCSPC noise distributions can be approximated by densities of the following form
\begin{equation}\label{dens}
f_\eta(x)=\left( \frac{\alpha\nu}{\alpha -\beta}e^{-\nu x}-\frac{\beta\tau}{\alpha -\beta}e^{-\tau x}\right) \1_{\{x>0\}}\;,
\end{equation}
with constraints $\alpha>\beta$, $\nu<\tau$, $\beta\tau/(\alpha\nu)\geq1$.
Figure \ref{fitted_density}  presents a dataset with 259,260 measurements from the noise distribution of a TCSPC instrument (independently from the fluorescence measurements) and  the corresponding estimated density having form
(\ref{dens}) obtained by least squares fitting. Even though the fit is not perfect, the estimated density captures the main features of the dataset.
Thus densities of the form (\ref{dens}) can be considered as a good approximative model of the noise distribution in the fluorescence setting.
In the general case of (\ref{dens}) we have
$$f_\eta^*(u)=\frac{\alpha\nu}{\alpha -\beta}\frac 1{\nu +iu}
-\frac{\beta\tau}{\alpha -\beta}\frac 1{\tau +iu}\;.$$
In the simulation study we will consider a noise distribution of the form (\ref{dens})  with  parameters  $\alpha= 2 , \beta= 1 , \nu= 1$, $\tau= 2$. In this case we get
\begin{equation}\label{exampleb} |f_\eta^*(u)|^2=\frac 4{(1+u^2)(4+u^2)}\qquad\text{ and }\qquad\Delta_\eta(m) = m+\frac 5{12}\pi^2m^3+\frac1{20}\pi^4m^5\;.\end{equation}

From the application viewpoint it is hence interesting to consider the class of noise distributions $\eta$ whose characteristic functions decrease in the \textit{ordinary smooth }way of order $\gamma$, denoted by $\eta \sim OS(\gamma)$, defined by
$c_0(1+u^2)^{-\gamma }\leq  |f_\eta^*(u)|^2 \leq
C_0(1+u^2)^{-\gamma }.$ Clearly, we find that $\Delta_\eta(m)=O(m^{2\gamma+1})$.

\subsection{Rates of convergence on Sobolev spaces}\label{rate}
In classical deconvolution  the regularity spaces used for the functions to estimate
are Sobolev spaces defined by
$${\mathcal C}(a,L)= \left\{g\in ({\mathbb L}^1\cap {\mathbb L}^2)({\mathbb R}),
\; \int (1+u^2)^a|g^*(u)|^2\rmd u \leq L\right\}\;.$$
If $f_Y$ belongs to ${\mathcal C}(a,L)$, then \begin{eqnarray*} \|f_Y-f_{Y,m}\|^2 &=& \int_{|u|\geq \pi m} |f_Y^*(u)|^2\rmd u =
\int_{|u|\geq \pi m} (1+u^2)^a|f_Y^*(u)|^2/(1+u^2)^a \rmd u\\ &\leq & (1+(\pi m)^2)^{-a} L\leq L(\pi m)^{-2a}\;.
\end{eqnarray*}

Therefore, if $f_Y\in{\mathcal C}(a,L)$ and $\eta\sim OS(\gamma)$, Proposition \ref{borne} implies that
${\mathbb E}(\|\bar f_m-f_Y\|^2) \leq C_1m^{-2a} +  C_2n^{-1}{m^{2\gamma+1}}.$
The optimization of this upper bound  provides the optimal
choice of $m$ by $m_{opt} =O(n^{1/(2a+2\gamma+1)})$ with
resulting rate ${\mathbb E}(\|\hat f_m-f_Y\|^2)\leq
O(n^{-2a/(2a+2\gamma +1)} )$. More formally, one can show the following result.

\begin{prop}\label{rateofdec}
Assume that the assumptions of Proposition \ref{borne} are satisfied and that $f_Y\in{\mathcal C}(a,L)$ and $\eta\sim OS(\gamma)$, then for $m_{opt} =O(n^{1/(2a+2\gamma+1)})$, we have
$${\mathbb E}(\|\bar f_{m_{opt}}-f_Y\|^2)\leq O(n^{-2a/(2a+2\gamma +1) } )\;.$$
\end{prop}

Obviously, in practice the optimal choice $m_{opt}$ is not feasible since $a$ is  and part of the constants involved
in the order are unknown. Therefore, another model selection device is required to choose a relevant $\bar f_m$ in the collection.

\subsection{Model selection}

The general method consists in finding a data driven penalty $\overline{{\rm pen}}(.)$ such
that the following model
\begin{equation}\label{mselect} \bar m=\arg\min_{m\in {\mathcal
M}_n} (\gamma_n^\dagger (\bar f_m) + \overline{{\rm pen}}(m))\end{equation}
achieves a bias-variance trade-off, where ${\mathcal M}_n$  has to be specified. In contrast to this general approach our result involves an additional $\ln(n)$-factor in the penalty compared to the variance order, which implies a loss with respect to the expected rate derived in Section \ref{rate}.

\begin{thm}\label{adaptconv} Assume that $f_Y$ is square integrable on ${\mathbb R}$, $\eta \sim OS(\gamma)$ and $w$ satisfies (\ref{hypw}) and (\ref{lip}).
Consider the estimator $\bar f_{\bar m}$  with model $\bar m$ defined by (\ref{mselect}) with penalty
\begin{equation}\label{pendec}
\overline{{\rm pen}}(m)= \kappa' \left(\int_0^1 w^2(u)\rmd u +  \kappa''c_w^2\ln(n)\right) \frac{
\Delta_\eta(m)}n\;,
\end{equation}
where $\kappa'$ and $ \kappa''$ are  numerical constants. Assume moreover that $\eta$ is ordinary smooth, i.e. $\eta\sim OS(\gamma)$, and that the model collection is described by ${\mathcal M}_n=\{m\in {\mathbb N}, \Delta_\eta(m)\leq n\}=\{1, \dots, m_n\}$. Then, there exist constants $\kappa', \kappa''$ such that
\begin{equation}\label{adaptdec} {\mathbb E}\left(\|\bar f_{\bar m}-f_Y\|^2\right) \leq C\left(
\inf_{m\in {\mathcal M}_n} \|f_Y- f_{Y,m}\|^2 + \overline{{\rm pen}}(m)\right) + C'\frac{\ln(n)}n\;,\end{equation}
where $C$ is a numerical constant and $C'$ depends on $c_w$ and the bounds on $w$.
\end{thm}
As previously, the numerical constants $\kappa'$ and $ \kappa''$  are calibrated via simulations. In practice,  to compute $\bar m$ by (\ref{mselect}), we approximate $\gamma_n^\dagger (\bar f_m)$  by $-\sum_{|j|\leq K_n} |\bar a_{m,j}|^2$, where the sum is truncated to $K_n$ of order $n$.

In the fluorescence set-up,  the noise distribution $f_\eta$ is generally unknown. However,  independent, large samples of the noise distribution are available. Hence one may still use the procedure proposed above by replacing $f_\eta^*$ with the estimate
$\hat f_\eta^*(u)=  \sum_{k=1}^n e^{-iu\eta_{-k}}/n,$
where $(\eta_{-k})_{1\leq k\leq M}$ denotes the independent noise sample.
In \cite{CL} the same substitution is considered for deconvolution methods. It is shown that for ordinary smooth noise
this leads to a risk bound exactly analogous to the one given in (\ref{adaptdec}). The main constraint given in \cite{CL}  is that $M\geq n^{1+\epsilon}$, for some $\epsilon>0$. As the noise samples provided in fluorescence have huge size, this condition is certainly fulfilled in our practical examples. In the following numerical study we consider the estimator with both the exact $f_\eta^*$ and an estimated $\hat f^*_\eta$.

\section{Numerical results for simulated and real data}\label{Simu}
In this section we first give details on the practical implementation of the estimation methods. Then a simulation study is conducted to test the performance of the methods in different settings. Finally, an application to a sample of  fluorescence data shows that the estimation method gives satisfying results on real measurements.

\subsection{Practical computation of estimators}
In the case of no additional noise, we apply the method described in Section \ref{firststep} with the  trigonometric basis [T].  To determine the best model $\hat m$ we compute $\gamma_n(m)+\text{pen}(m)$  for all $m=1,\dots,[n/2]-1$. This is computationally easy as the following recursive relation can be used. We have
$\gamma_n(0)+\text{pen}(0) = -\hat a_0^2 +  \kappa W/n,$ $\gamma_n(1)+\text{pen}(1) = -\hat a_0^2 -\hat a_{1,1}^2 -\hat a_{1,2}^2 + \kappa 3W/n$ and
$\gamma_n(m+1)+\text{pen}(m+1) = \gamma_n(m)+\text{pen}(m) - \hat a_{m+1,1}^2-\hat a_{m+1,2}^2 +2\kappa W/n$, for all $m\geq1$,
where $W=\int_0^1w^2(u)\rmd u$.
The coefficients are given by (\ref{estimbias}).
Then $\hat m$ is the value where $\gamma_n(m)+\text{pen}(m)$ achieves its minimum. Finally, the estimator of $f$ is given by $\hat f_{\hat m} = \sum_{\lambda\in\Lambda_{\hat m}}\hat a_\lambda\varphi_\lambda$.

In the case of additional noise, we use the estimator proposed in Section \ref{secondstep} based on the sinc basis. Its computation is more intensive as no similar recursive relation holds. First one has to compute the coefficients
$\bar a_{m,j}$ defined in (\ref{explicit}). For $j\geq0$ they can be approximated as follows
\begin{align*}
\bar a_{m,j}
&= \frac1{2\pi} \int \varphi_{m,j}^*(-u) \frac {\widehat {f^*_X}(u) }{f_\eta^*(u)}\rmd u
= (-1)^j\frac {\sqrt m}2 \int_0^2 e^{i\pi j v} \frac {\widehat {f^*_X}(\pi m(v-1))}{f_\eta^*(\pi m(v-1))}\rmd v\\
&\approx (-1)^j  \frac {\sqrt m}{T}\sum_{t=0}^{T-1} e^{i2\pi jt/T} \frac {\widehat {f^*_X}(\pi m(\frac{2 t}T-1))}{f_\eta^*(\pi m(2t/T-1))}
= (-1)^j{\sqrt m}(\text{IFFT}(H))_j = \breve a_{m,j}
\;,
\end{align*}
where IFFT(H) is the inverse fast Fourier transform of the $T$-vector $H$ whose $t$-th entry equals $\widehat {f^*_X}(\pi m({2 t}/T-1))/{f_\eta^*(\pi m(2t/T-1))}$. Similarly, for $j<0$ the coefficients $\bar a_{m,j}$ are approximated by $\breve a_{m,j} = (-1)^j{\sqrt m}(\text{IFFT}(\overline{H}))_j$.

The integral $\Delta_\eta(m)$ appearing in the penalty term $\overline{{\rm pen}}(m)$ defined in (\ref{pendec}) is explicitly  known  if $f_\eta$ is known (see Section \ref{OS}).
In the case when we only have an estimator $\hat f_\eta$, $\Delta_\eta(m)$ can be approximated by a Riemann sum of the form $({m}/{S}) \sum_{s=0}^S{|\hat f^*_\eta(-\pi m(1-\frac{2s}S))|^{-2}}$.
Then the best model $\bar m$ is selected as the point of minimum of the criterion given in (\ref{mselect}).
Finally, we obtain the estimator $\bar f_{\bar m} = \sum_{j=-T}^T\breve a_{\bar m, j}\varphi_{\bar m,j}$ with the sinc functions $\varphi_{m,j}$ defined in (\ref{sinc_fn}).

\begin{center}
\begin{figure}
\begin{tabular}{cccc}
Gamma & Exponential & Pareto & Weibull \\
\includegraphics[scale=0.12]{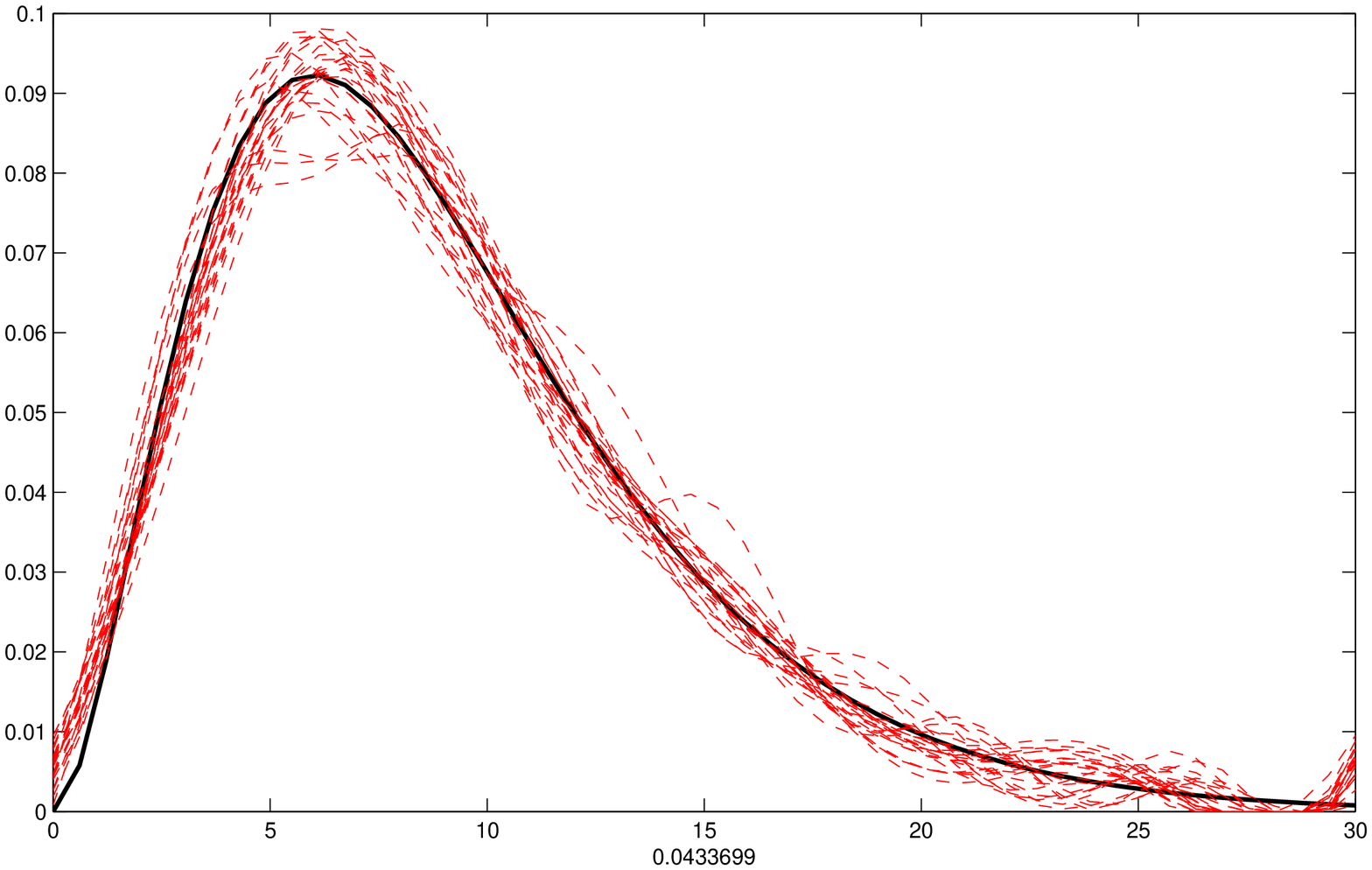}
& \includegraphics[scale=0.12]{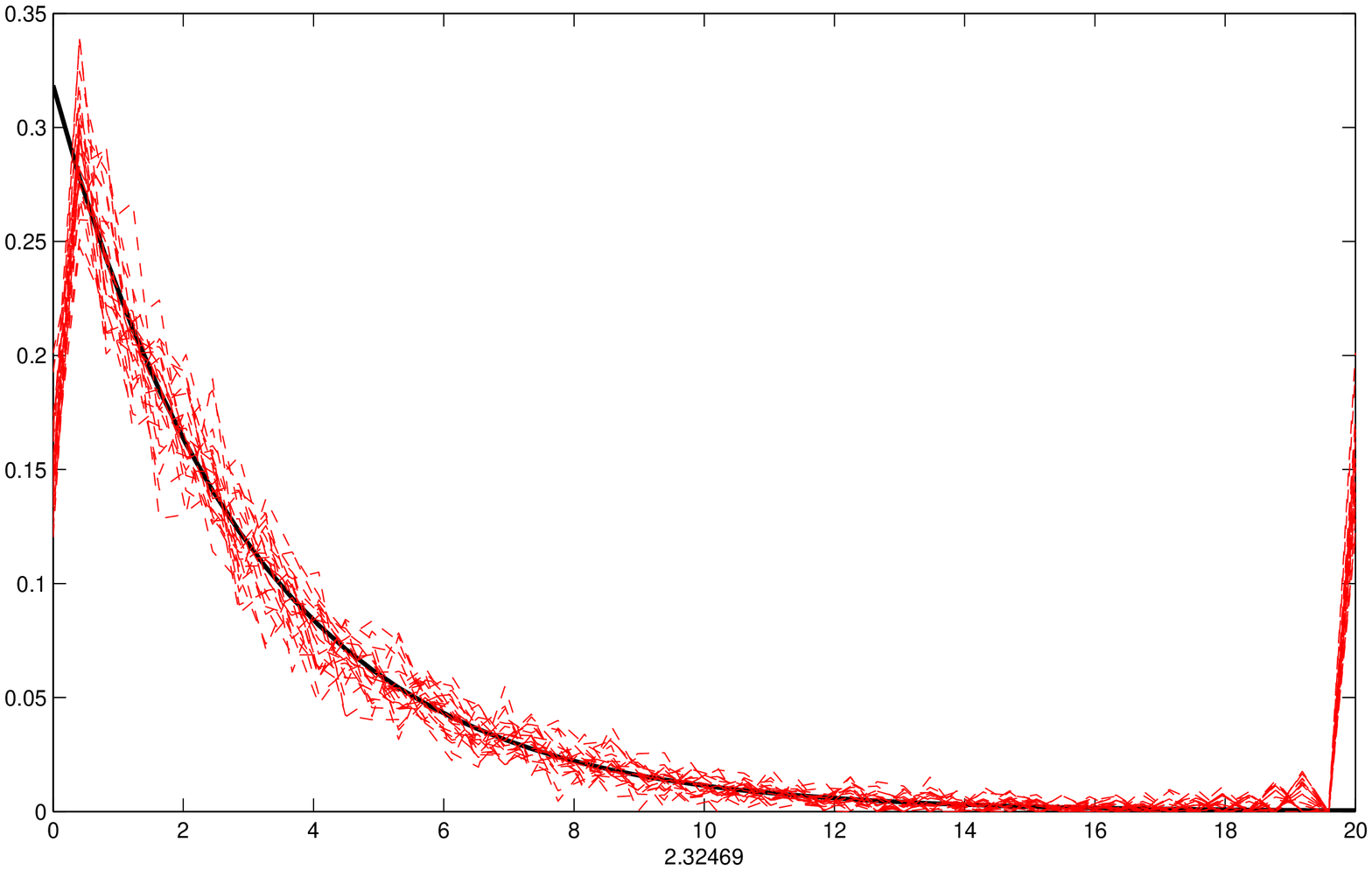}
& \includegraphics[scale=0.12]{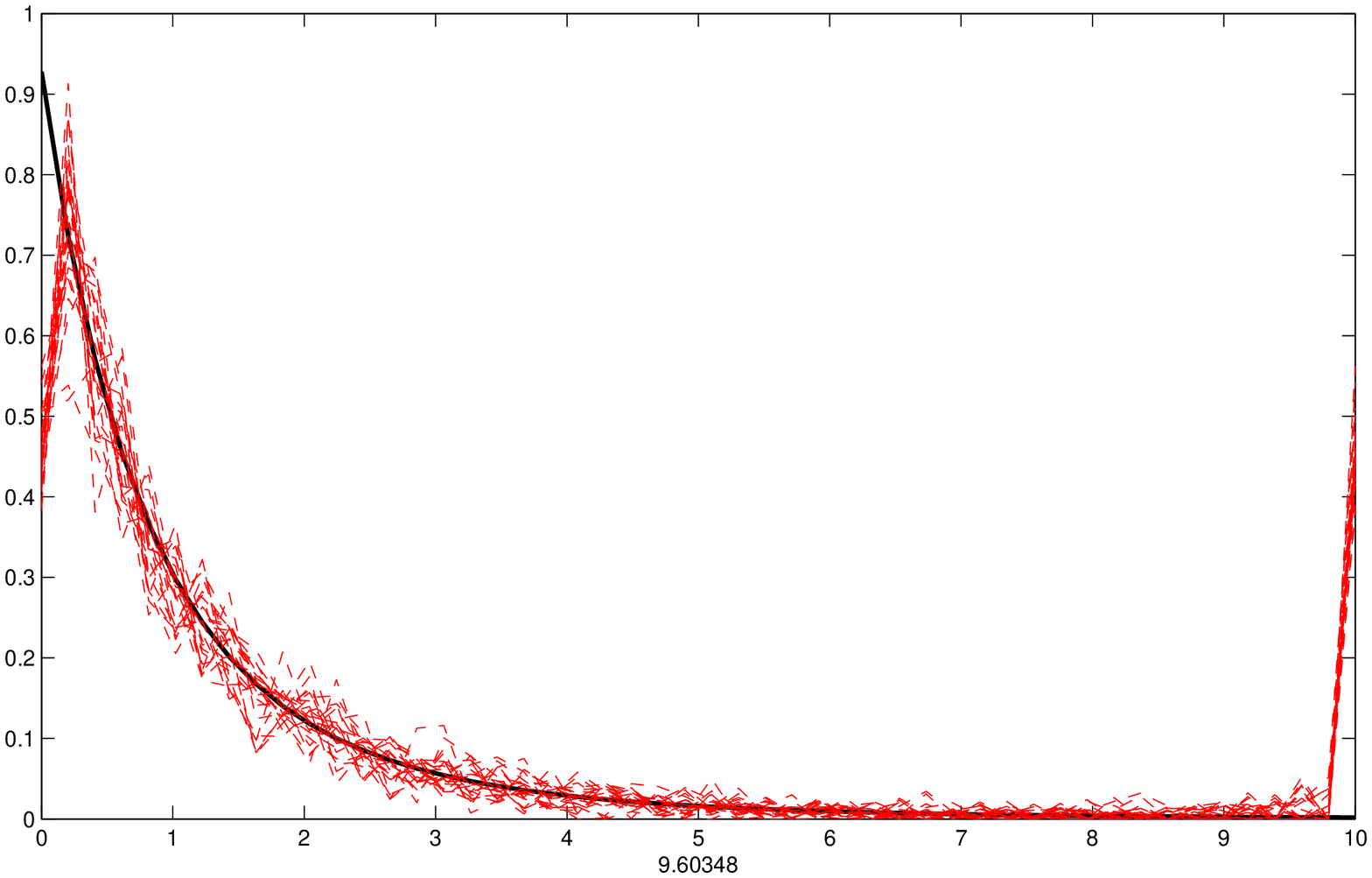}
&\includegraphics[scale=0.12]{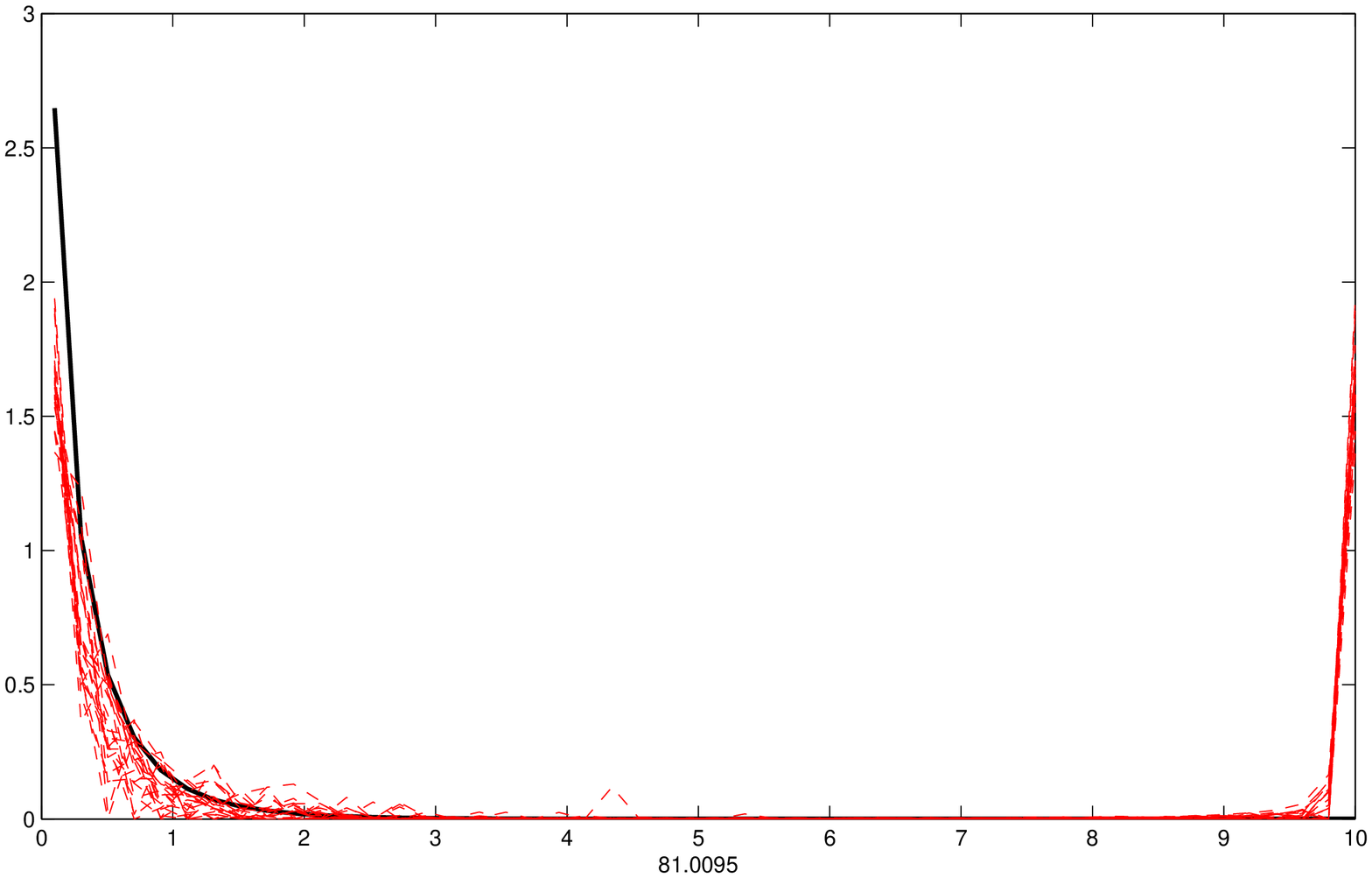}\\
 0.0434&2.325&9.6035&81.010\\
\multicolumn{4}{c}{Estimation with $\mu=0.01$, $n=1000$}\\
~&&&\\
\includegraphics[scale=0.12]{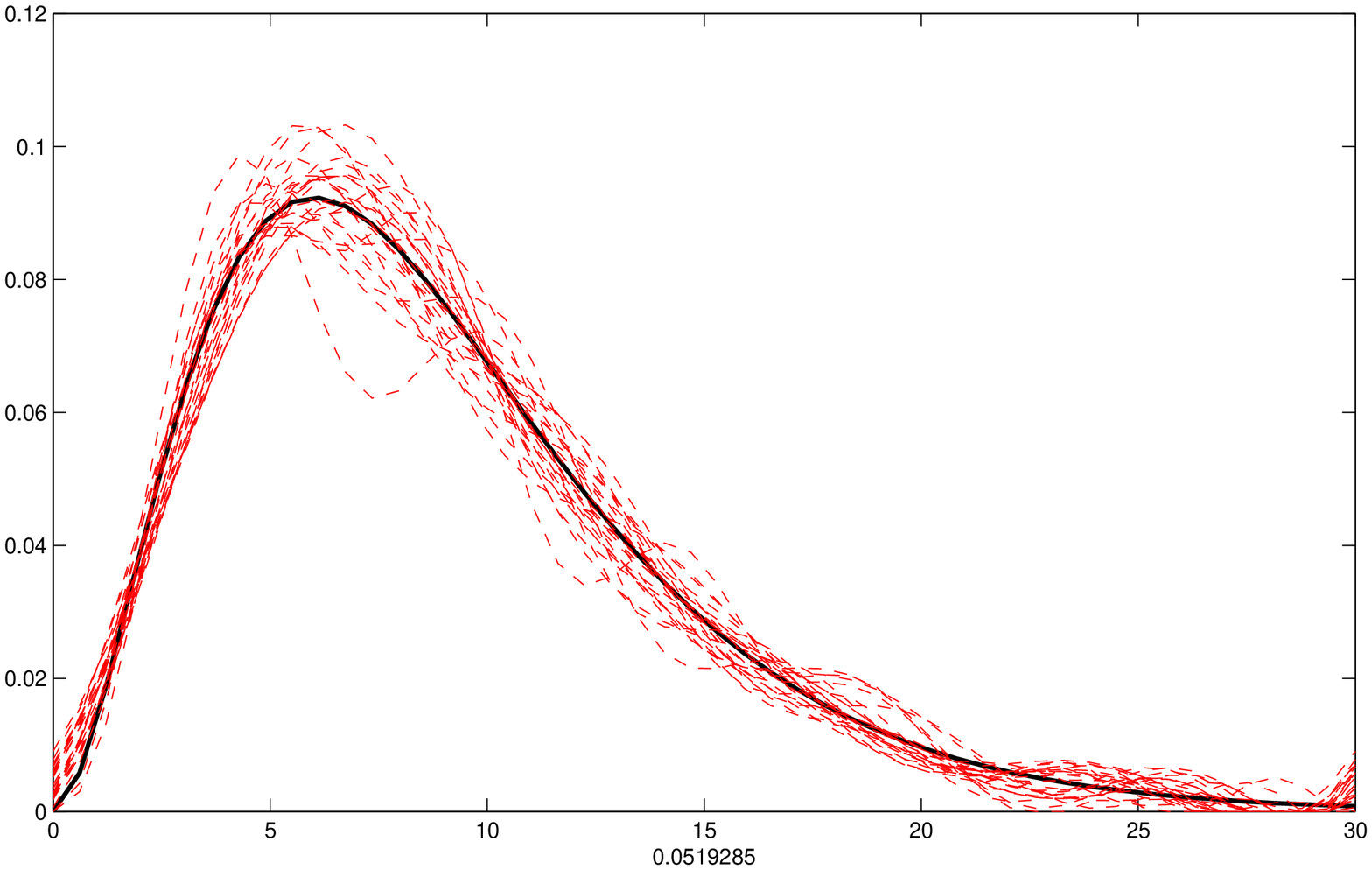}
& \includegraphics[scale=0.12]{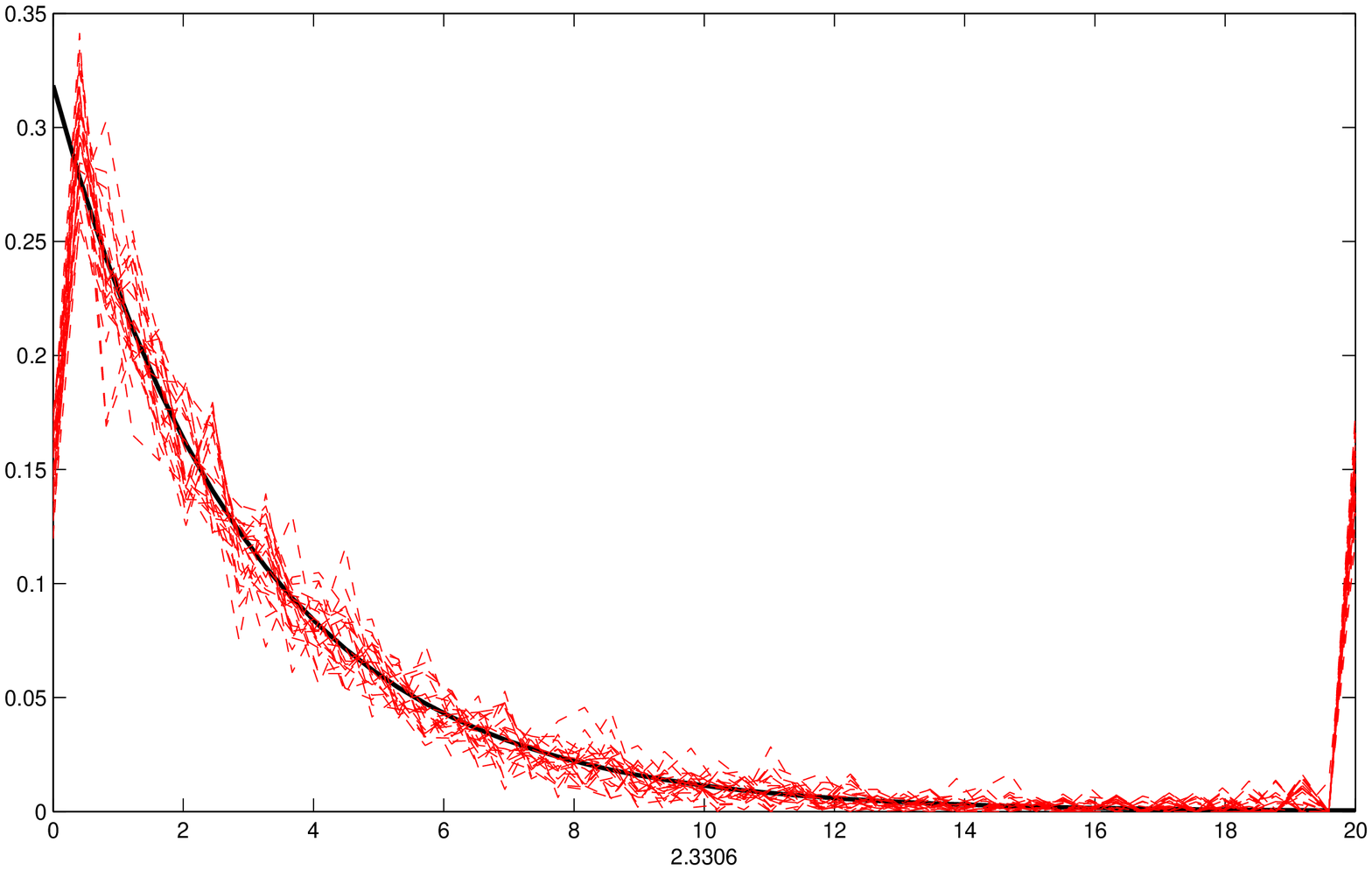}
& \includegraphics[scale=0.12]{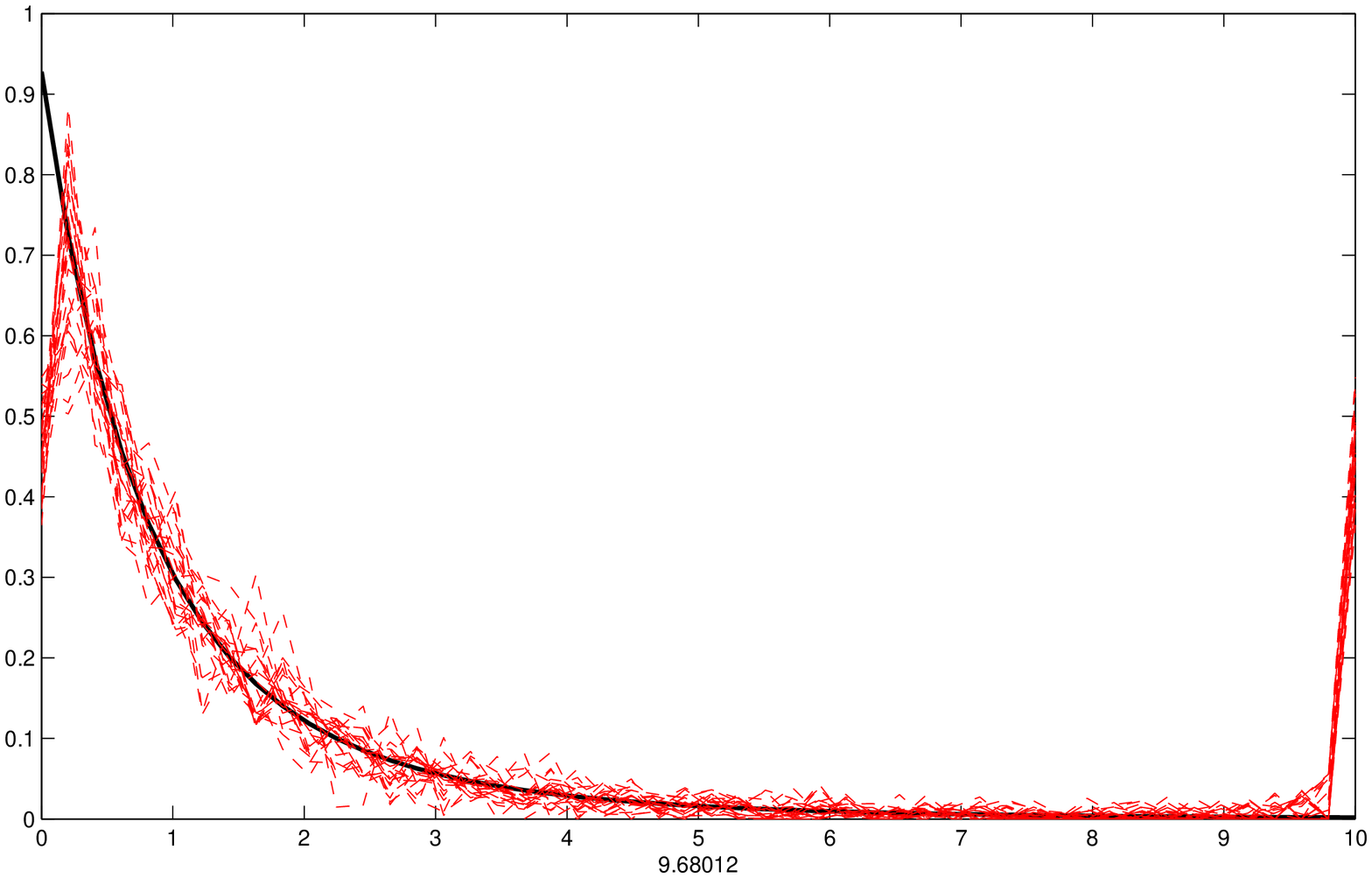}
&\includegraphics[scale=0.12]{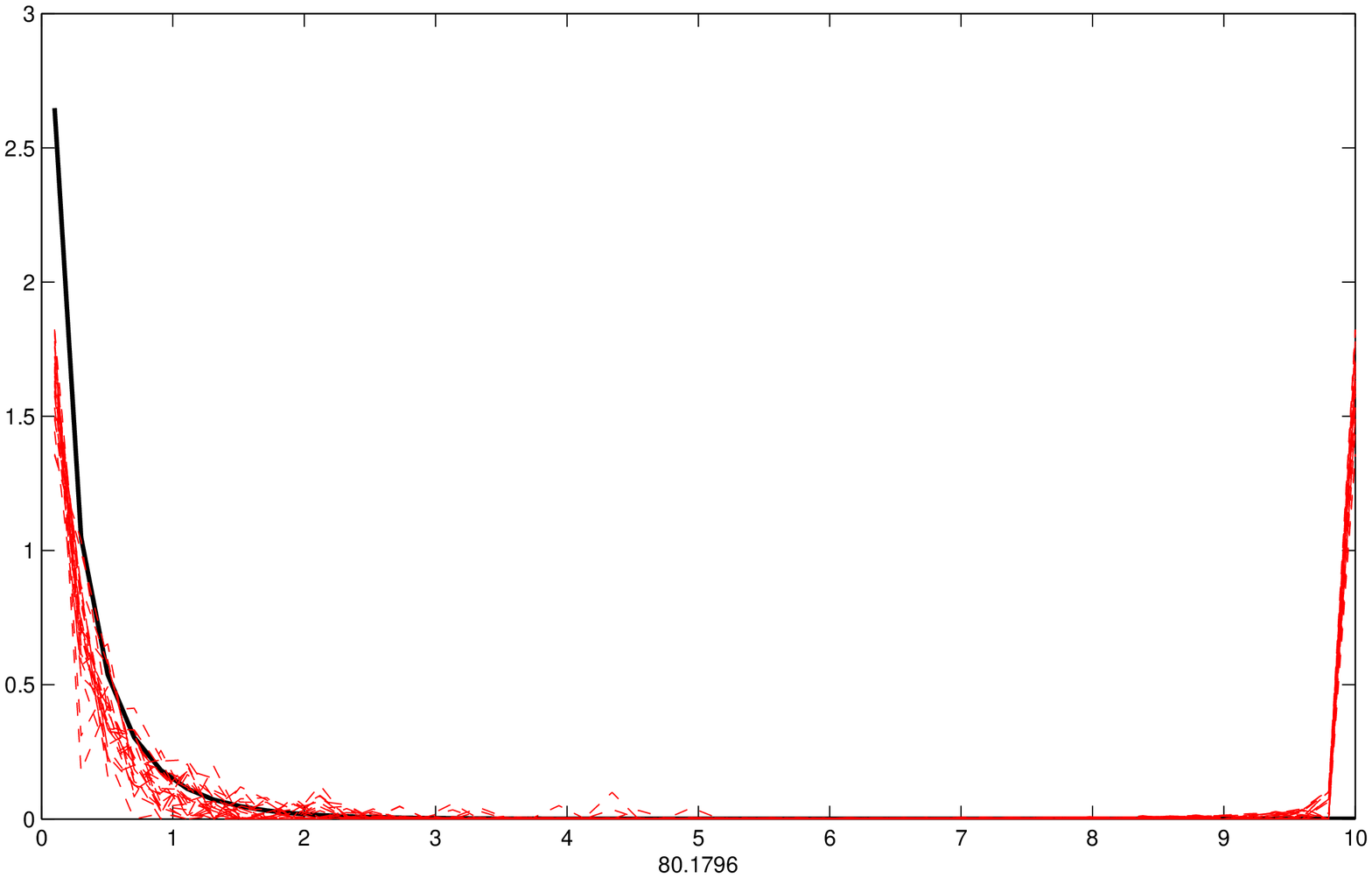}\\
 0.0519&2.3306&9.6801&80.1796\\
\multicolumn{4}{c}{Estimation with $\mu=0.5$, $n=1000$}\\
~&&&\\
\includegraphics[scale=0.12]{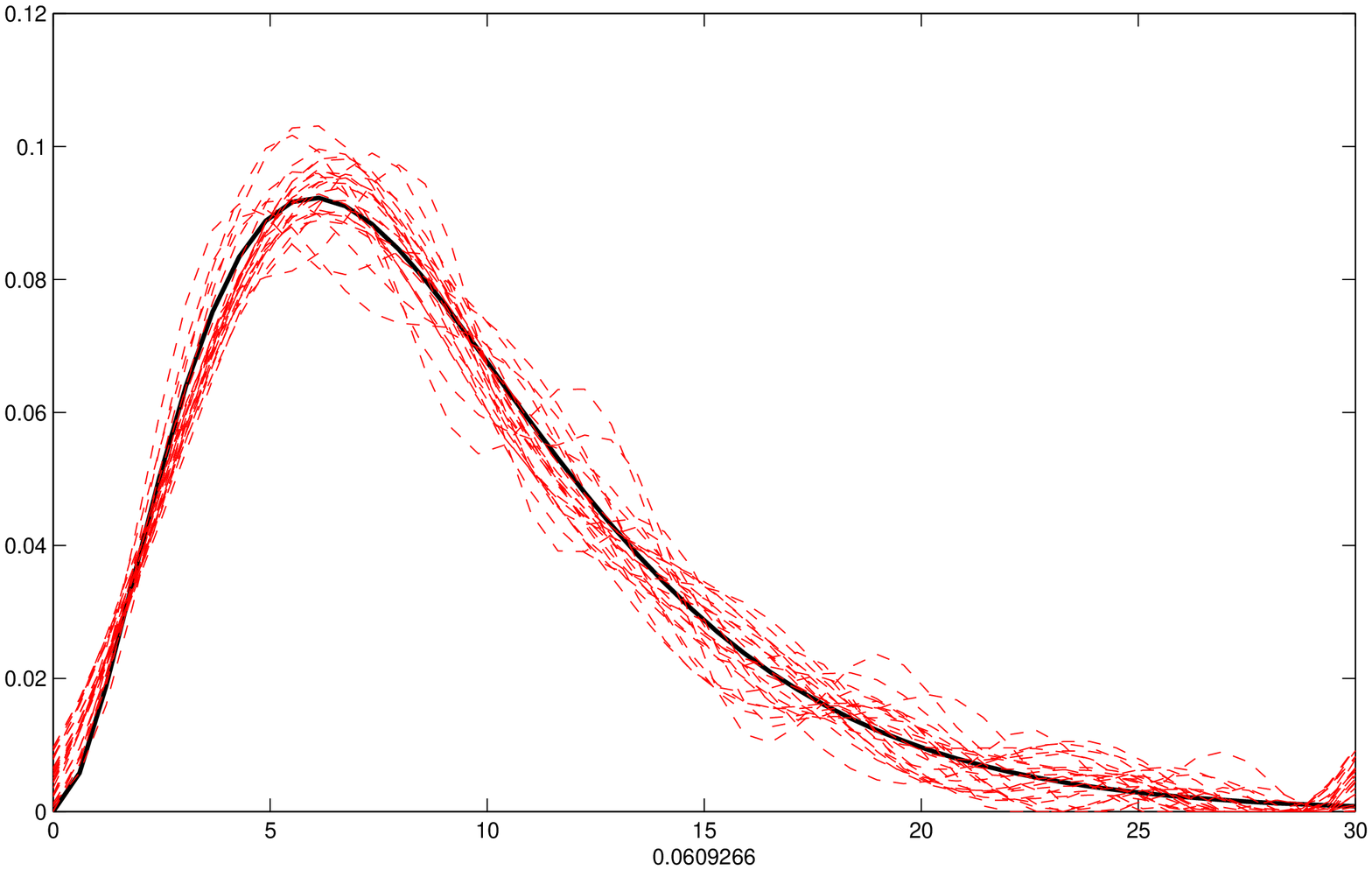}
& \includegraphics[scale=0.12]{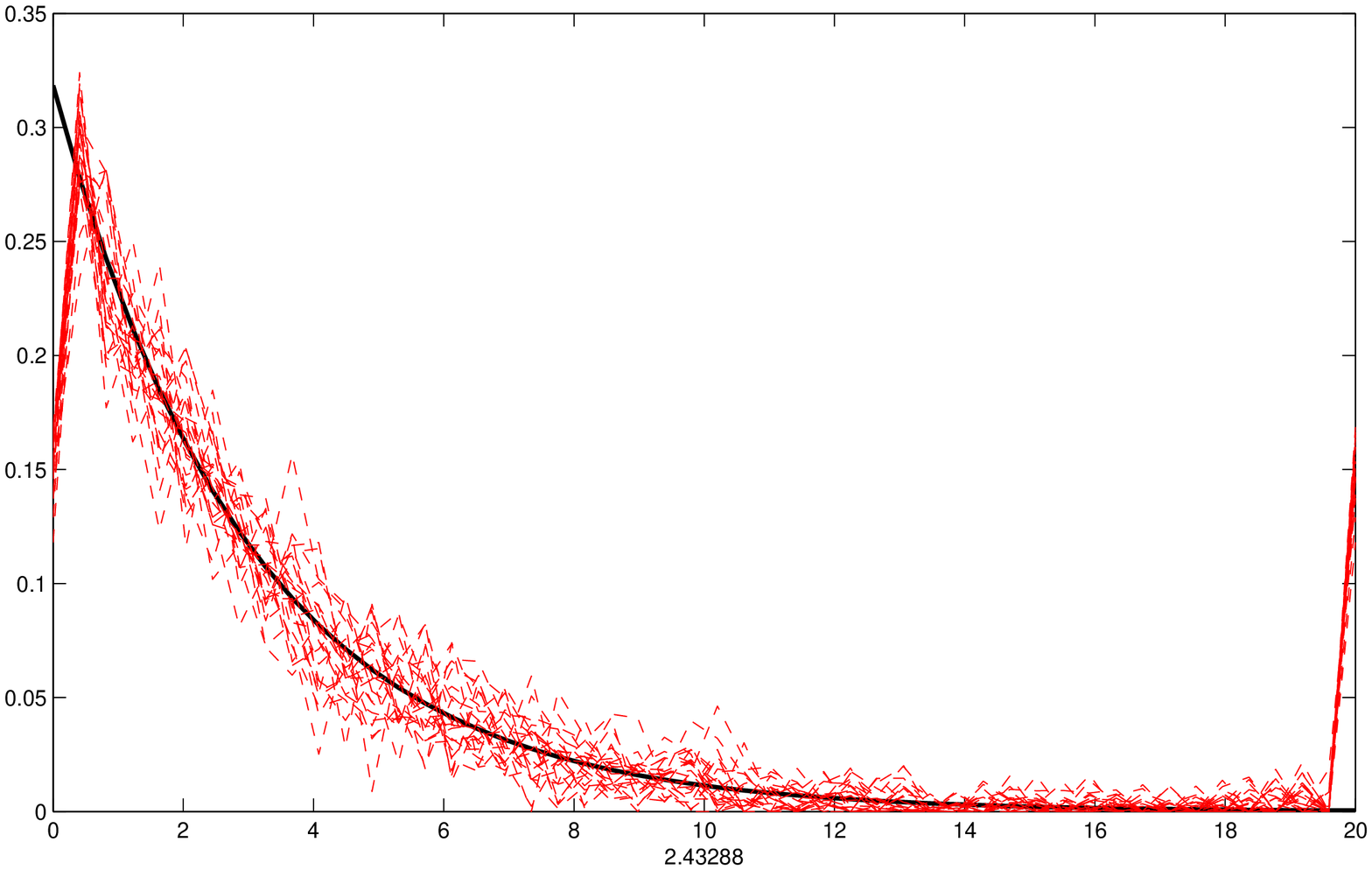}
& \includegraphics[scale=0.12]{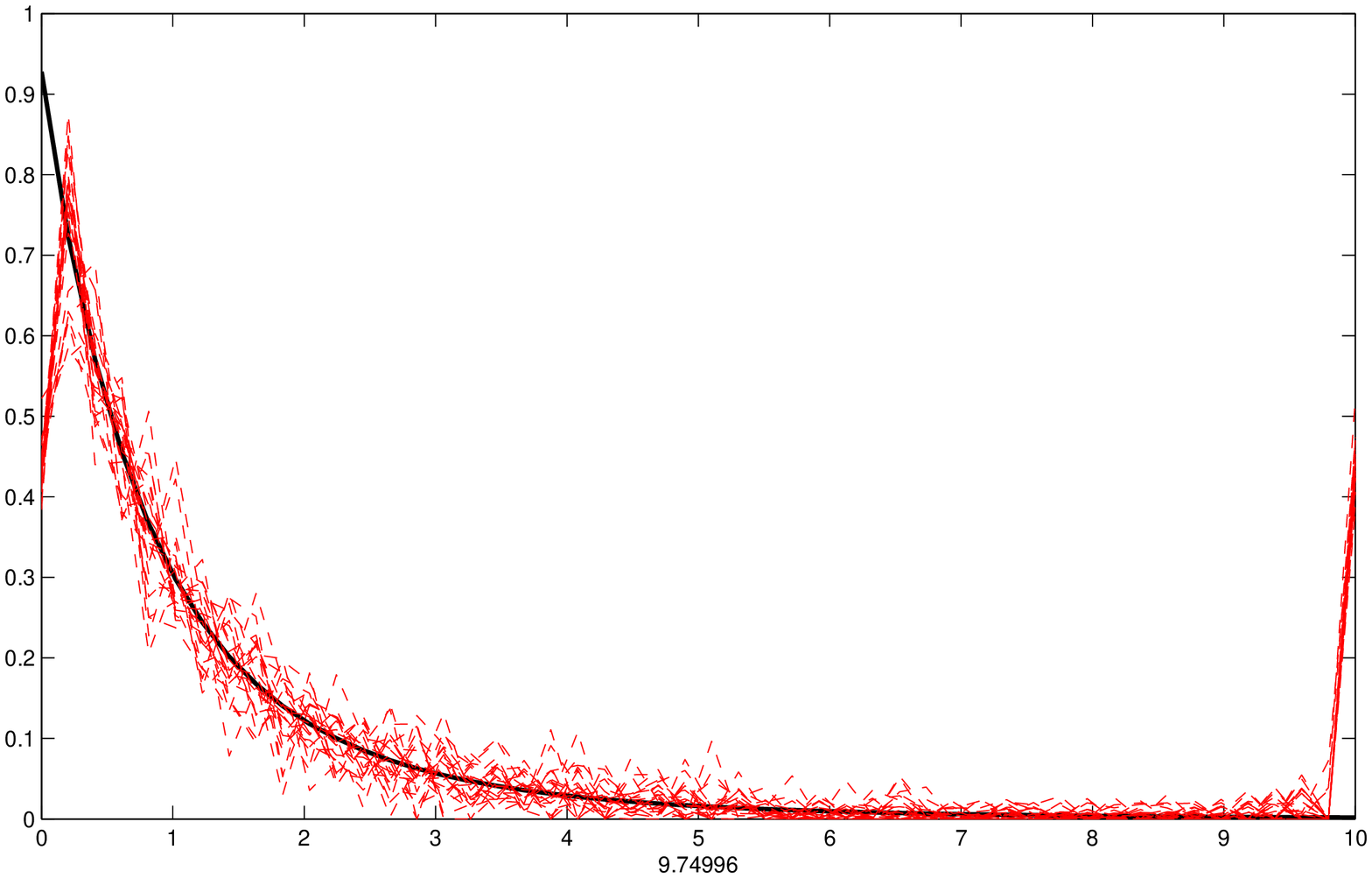}
&\includegraphics[scale=0.12]{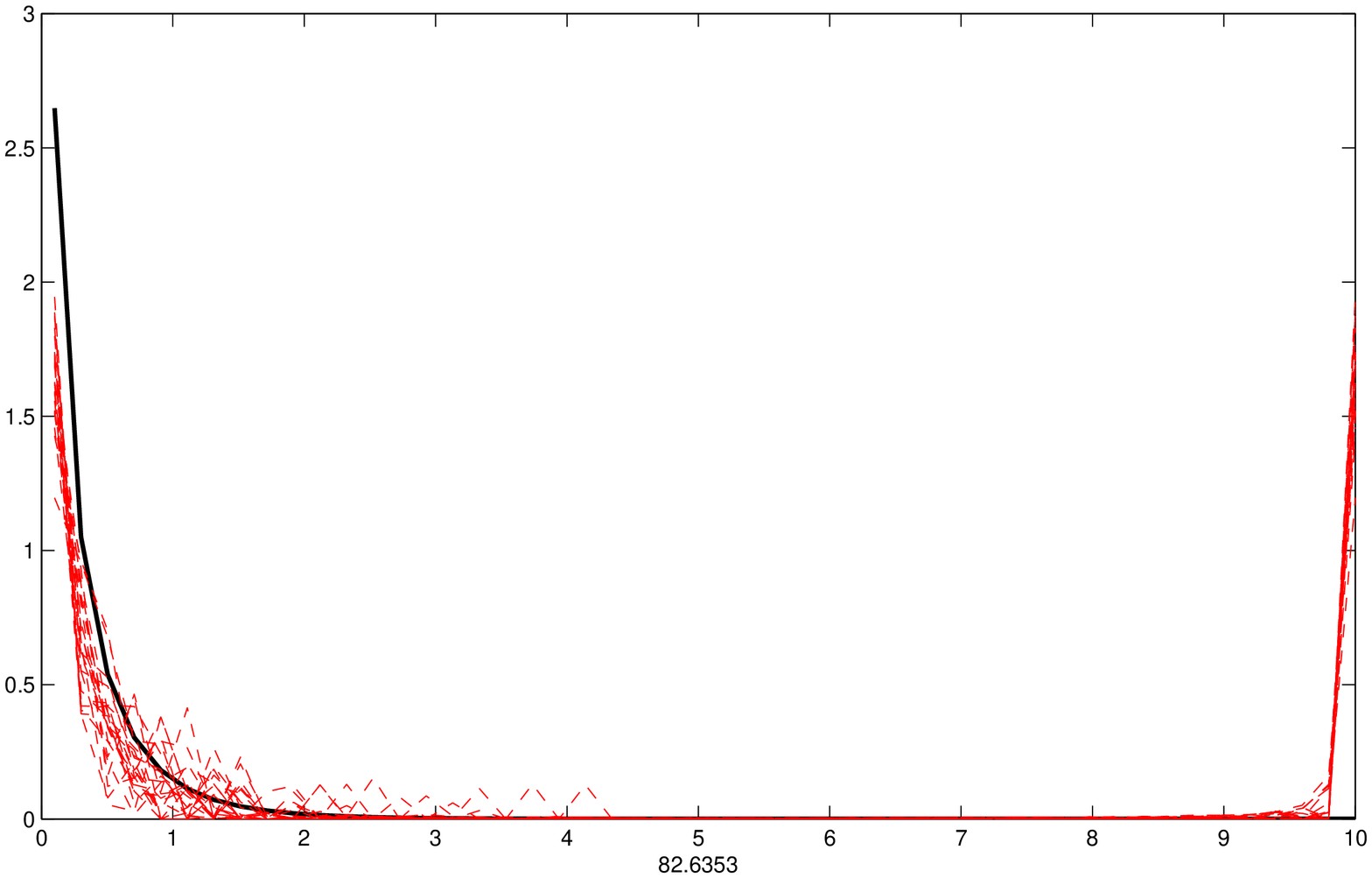}\\
 0.0610 &2.4329&9.7500&82.6353\\
\multicolumn{4}{c}{Estimation with $\mu=2$, $n=1000$}\\
\end{tabular}
\caption{ True density and 25 estimated curves without measurement errors. Estimation with the trigonometric basis for different levels of the pile-up effect.  Numbers below the figures are the MISE. }\label{planche_nonoise}
\end{figure}
\end{center}

Figure \ref{planche_nonoise} and \ref{planche_noisy} present the visual summary of our simulation results.
We implemented the estimation methods when $f_Y$ has one of the following pdfs.
\begin{enumerate}\addtolength{\itemsep}{-.7\baselineskip}
\item a Gamma(3, 3) p.d.f, $1/(2!3^3)x^2\exp(-x/3)\1_{x>0}$, to have a benchmark with a smooth distribution,
\item an exponential  pdf, $(1/3)\exp(-x/3)\1_{x>0}$,
\item a Pareto(1/4, 1, 0)    pdf $(1+x/4)^{-5}\1_{x>0}$,
\item a Weibull(1/4, 3/4)   pdf $(3/4)(1/4)^{-3/4} x^{-1/4}\exp(-(4x)^{3/4})\1_{x>0}$.
\end{enumerate}
The last two densities are inspired by chemical results about fluorescence phenomena given in \cite{BBVa,BBVb}.

\subsection{Simulation study}
\begin{center}
\begin{figure}
\begin{tabular}{cccc}
Gamma & Exponential & Pareto & Weibull \\
\includegraphics[scale=0.15]{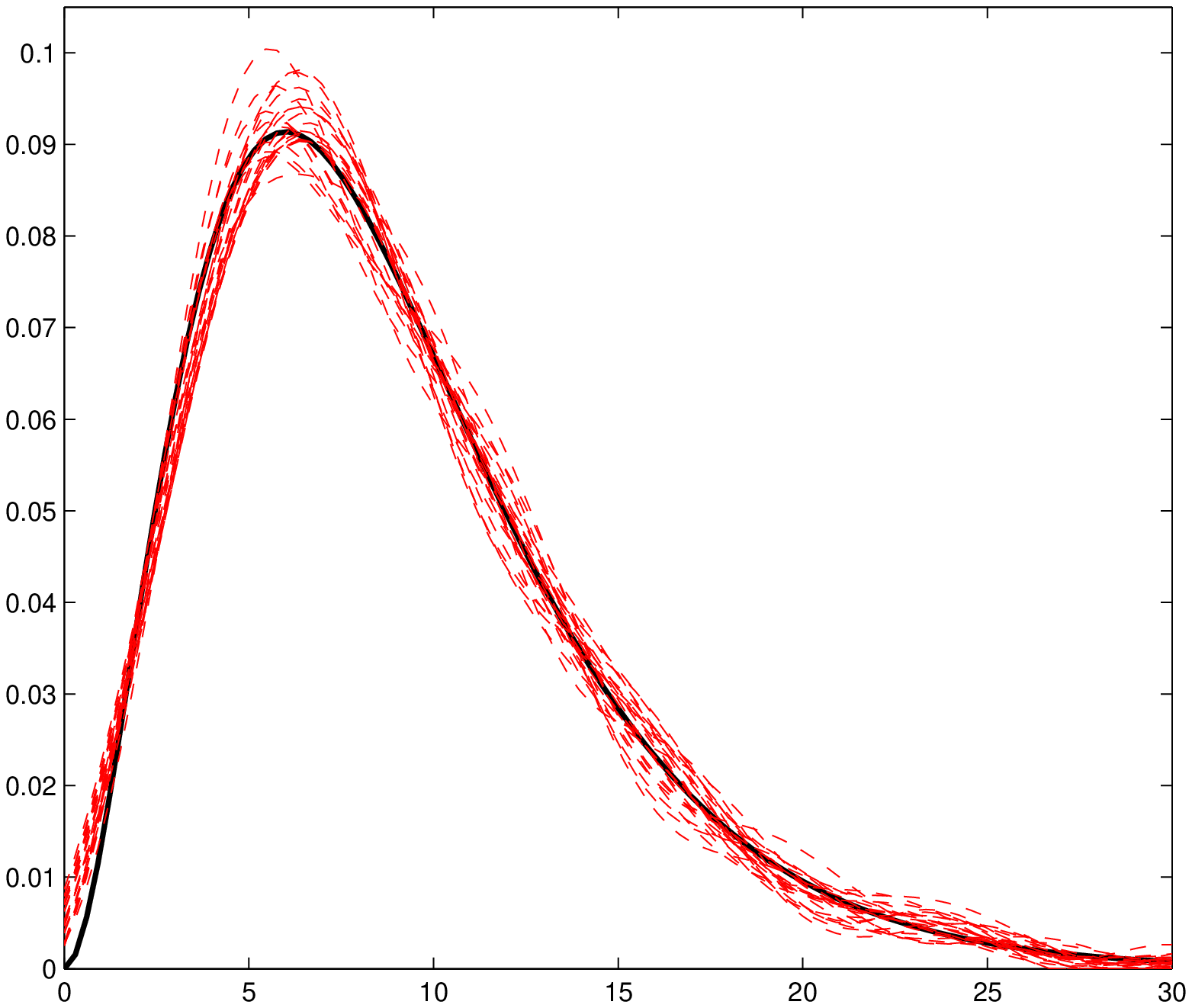}
& \includegraphics[scale=0.15]{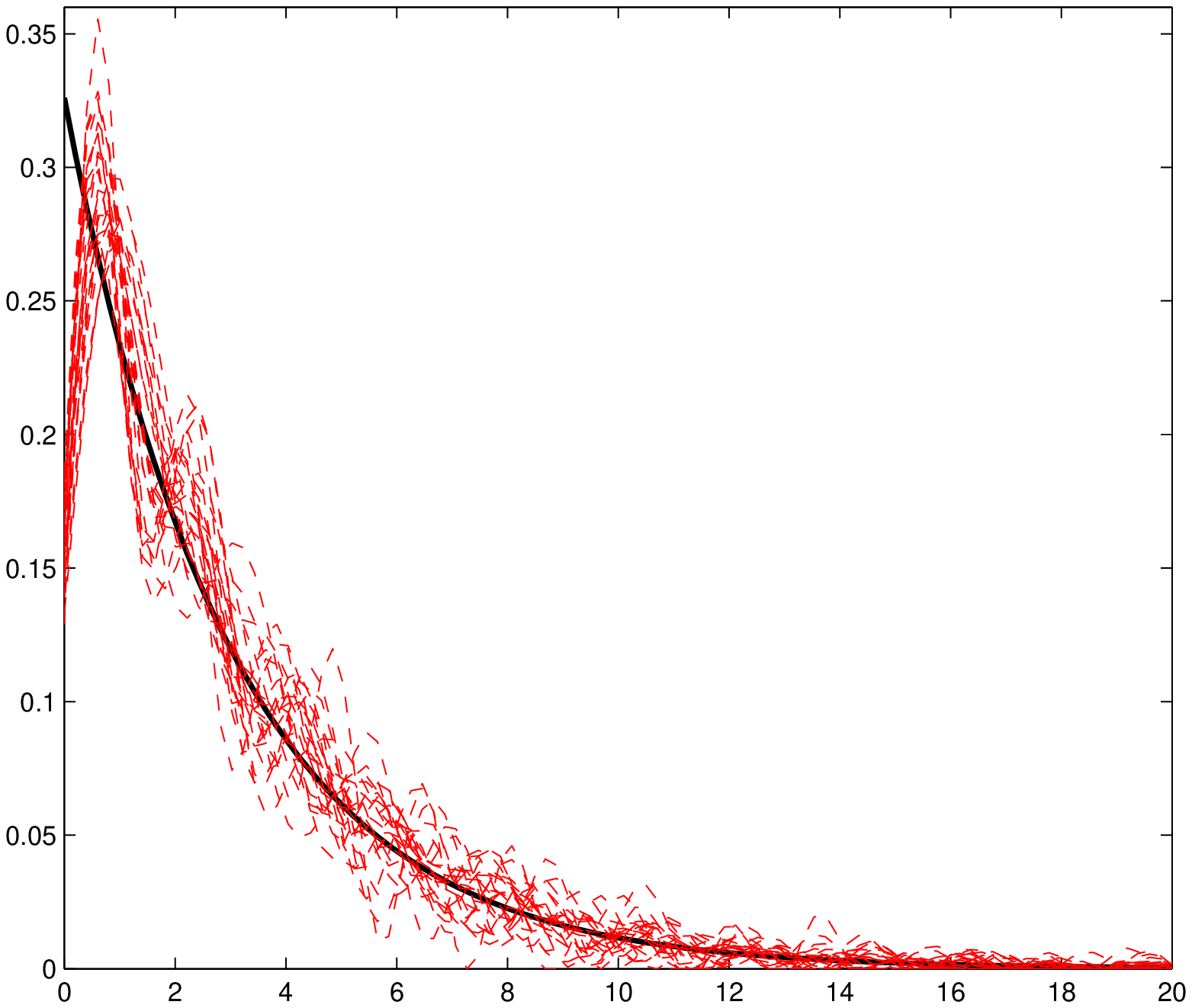}
& \includegraphics[scale=0.15]{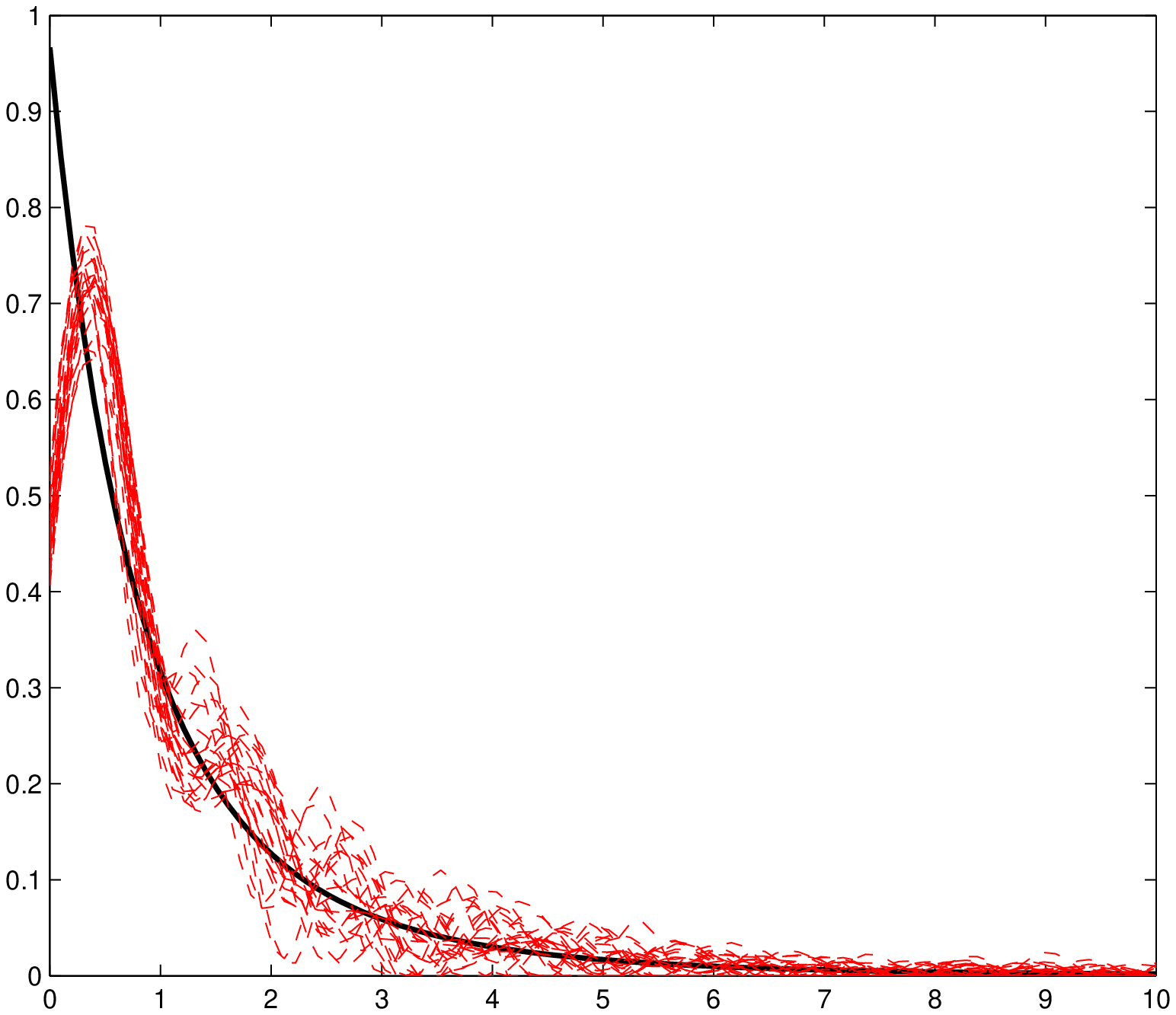}
&\includegraphics[scale=0.15]{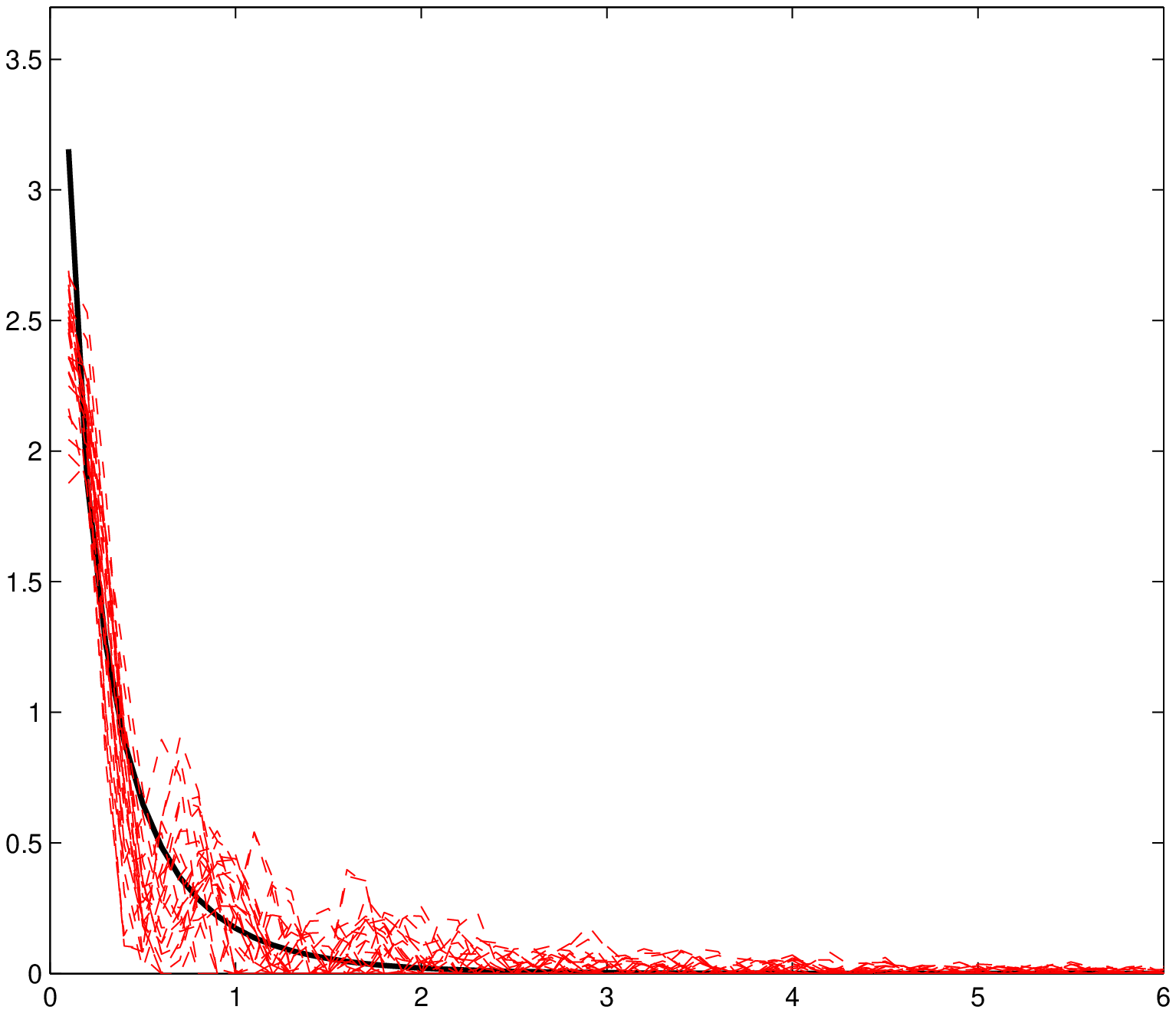}\\
3.36 (0.49)&14.2 (2.1)&22.4 (1.8)& 43.1 (5.9)\\
\multicolumn{4}{c}{Estimation with $\sigma=0.7$, $\mu=0.01$, $n=2000$}\\
~&&&\\
\includegraphics[scale=0.15]{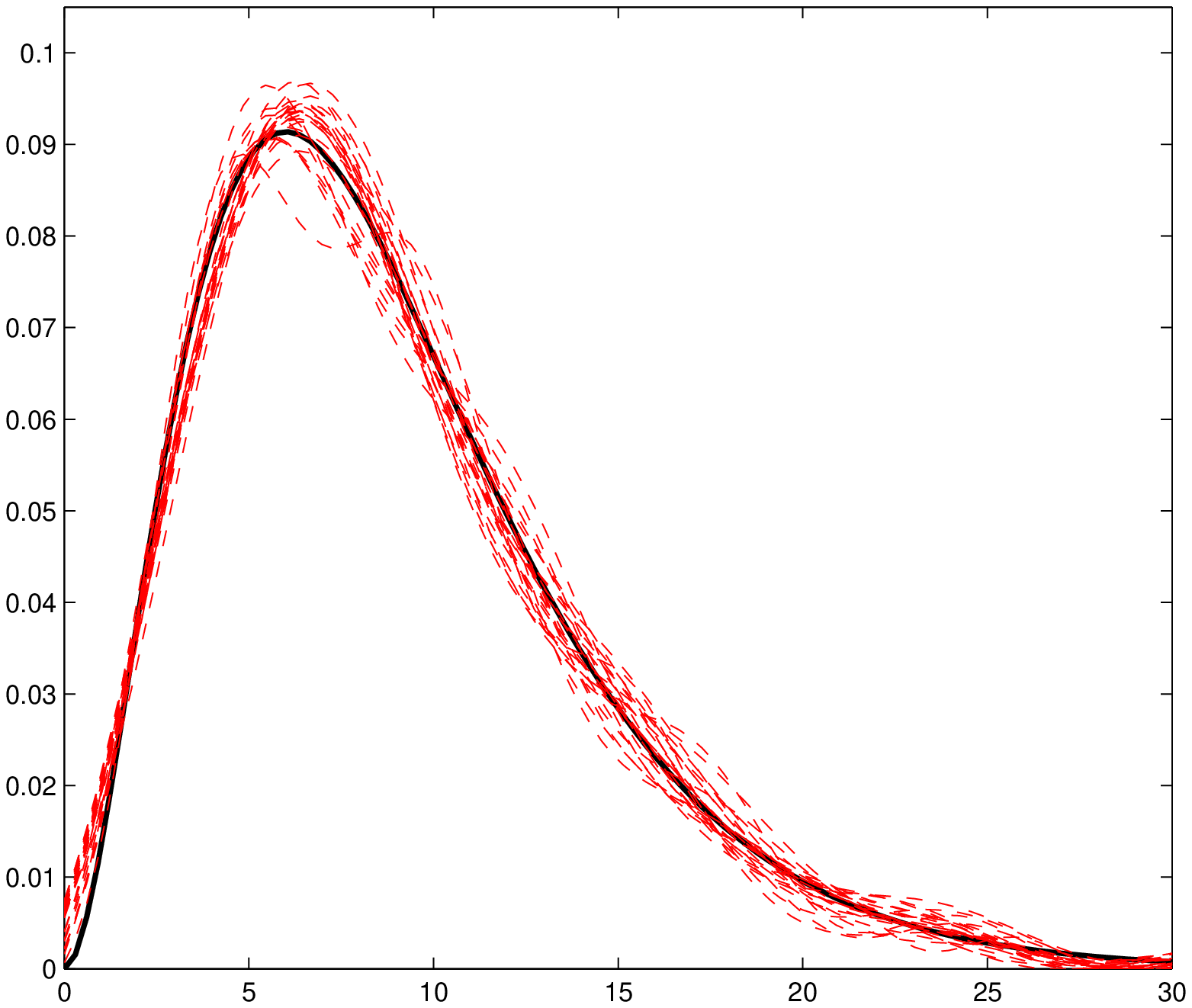}
& \includegraphics[scale=0.15]{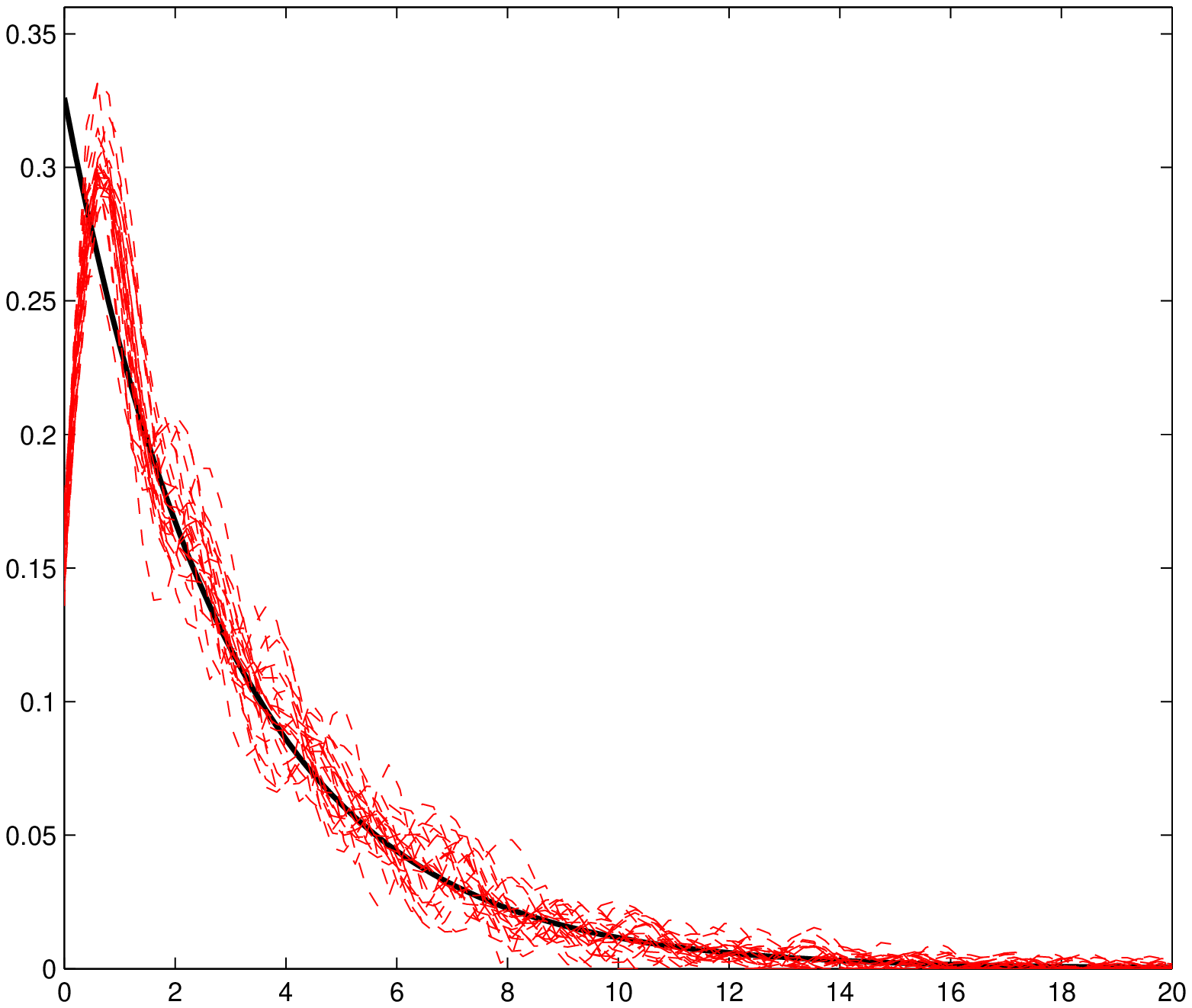}
& \includegraphics[scale=0.15]{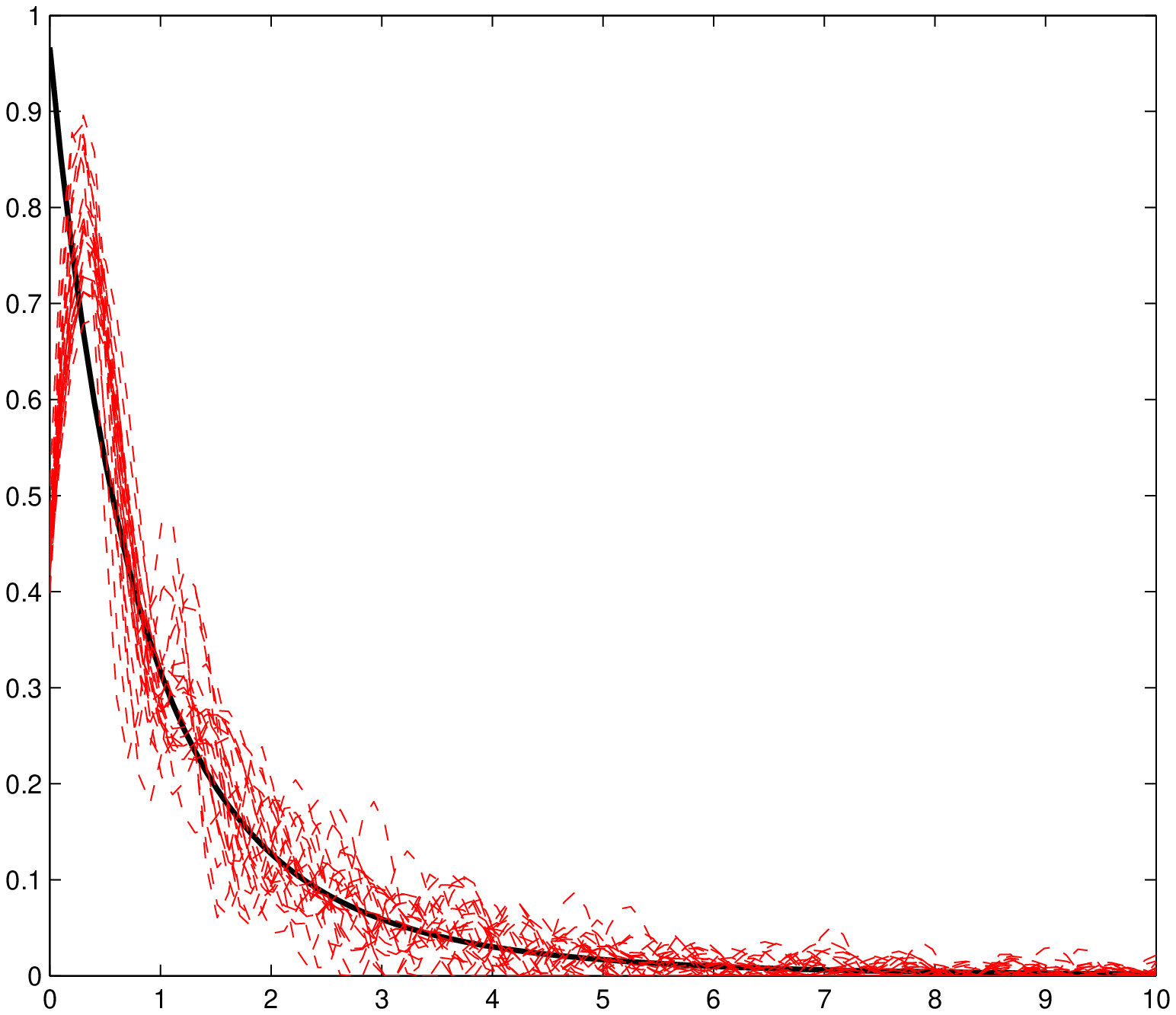}
&\includegraphics[scale=0.15]{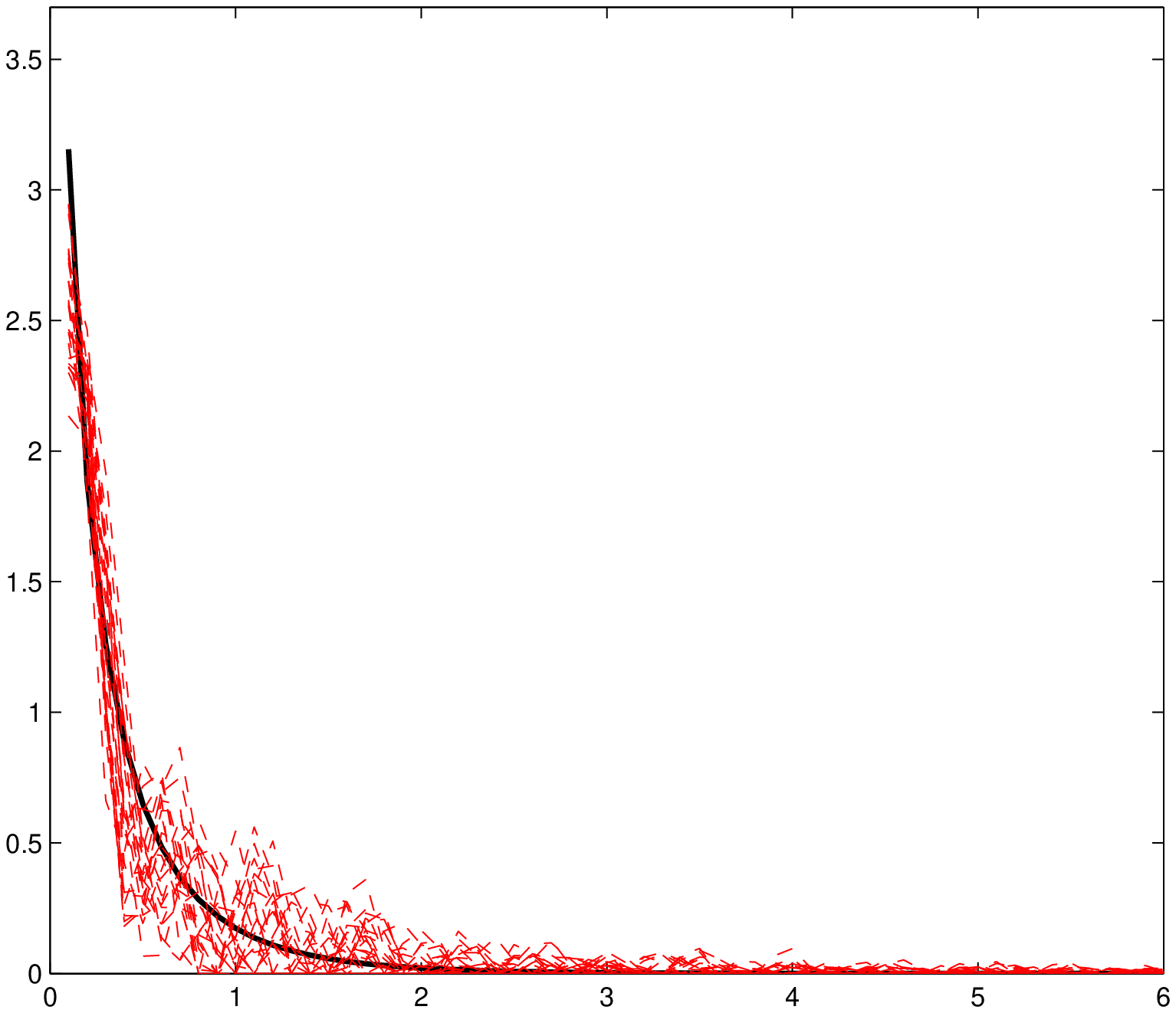}\\
3.36 (0.57)&15.2 (1.6)&27.1 (3.6)&48.9 (6.3)\\
\multicolumn{4}{c}{Estimation with $\sigma=0.5$, $\mu=1$, $n=2000$}\\
~&&&\\
\includegraphics[scale=0.15]{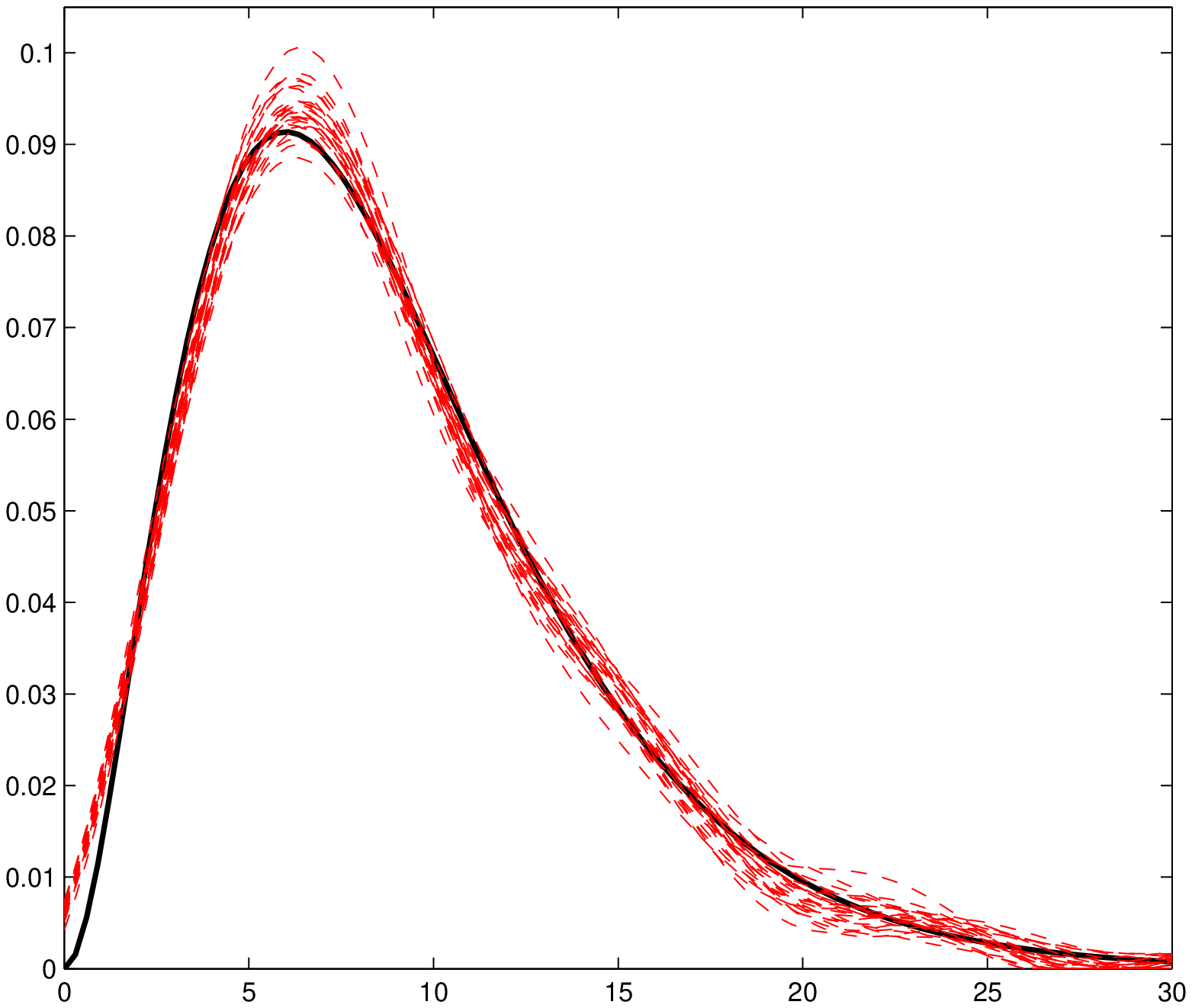}
& \includegraphics[scale=0.15]{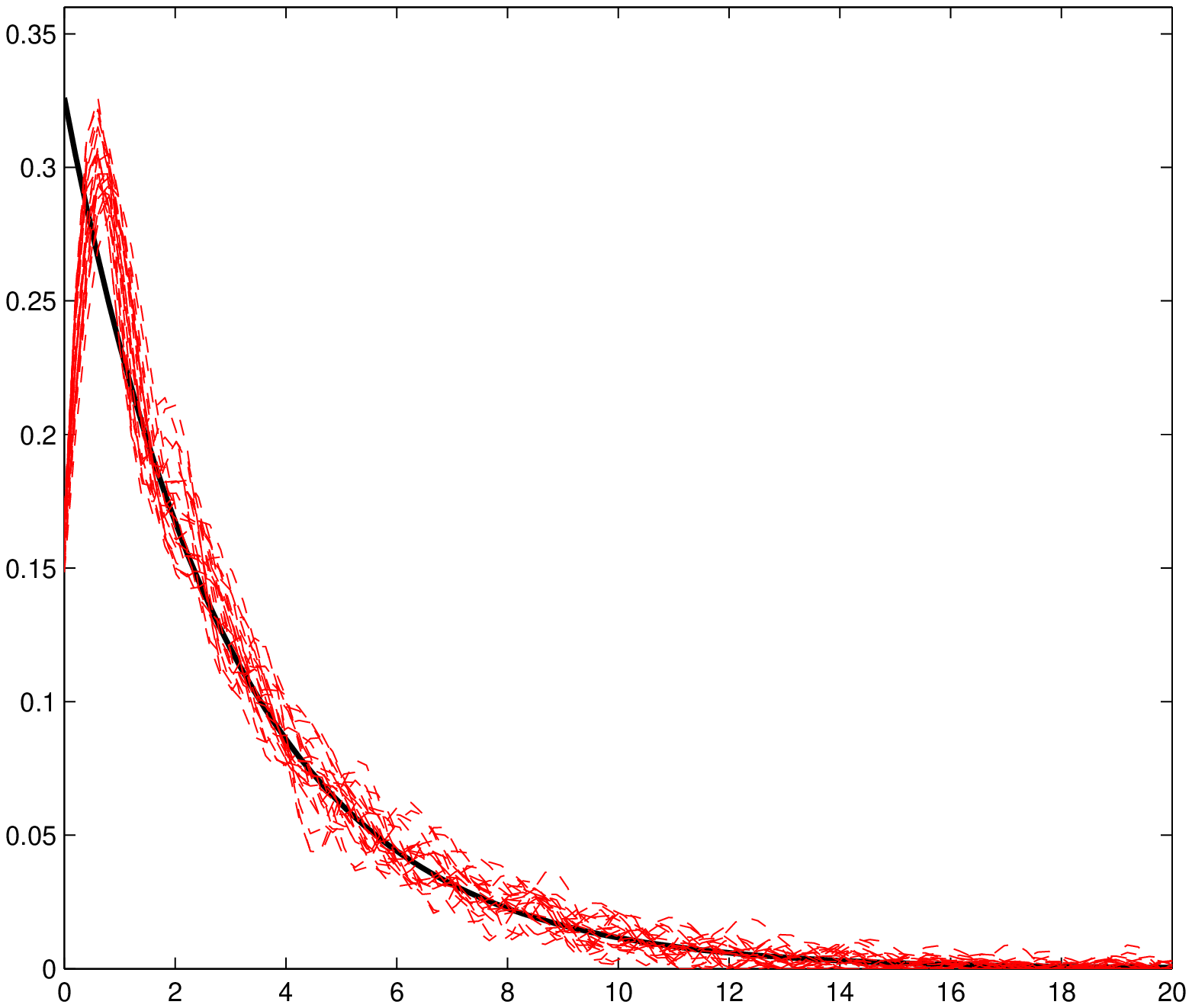}
& \includegraphics[scale=0.15]{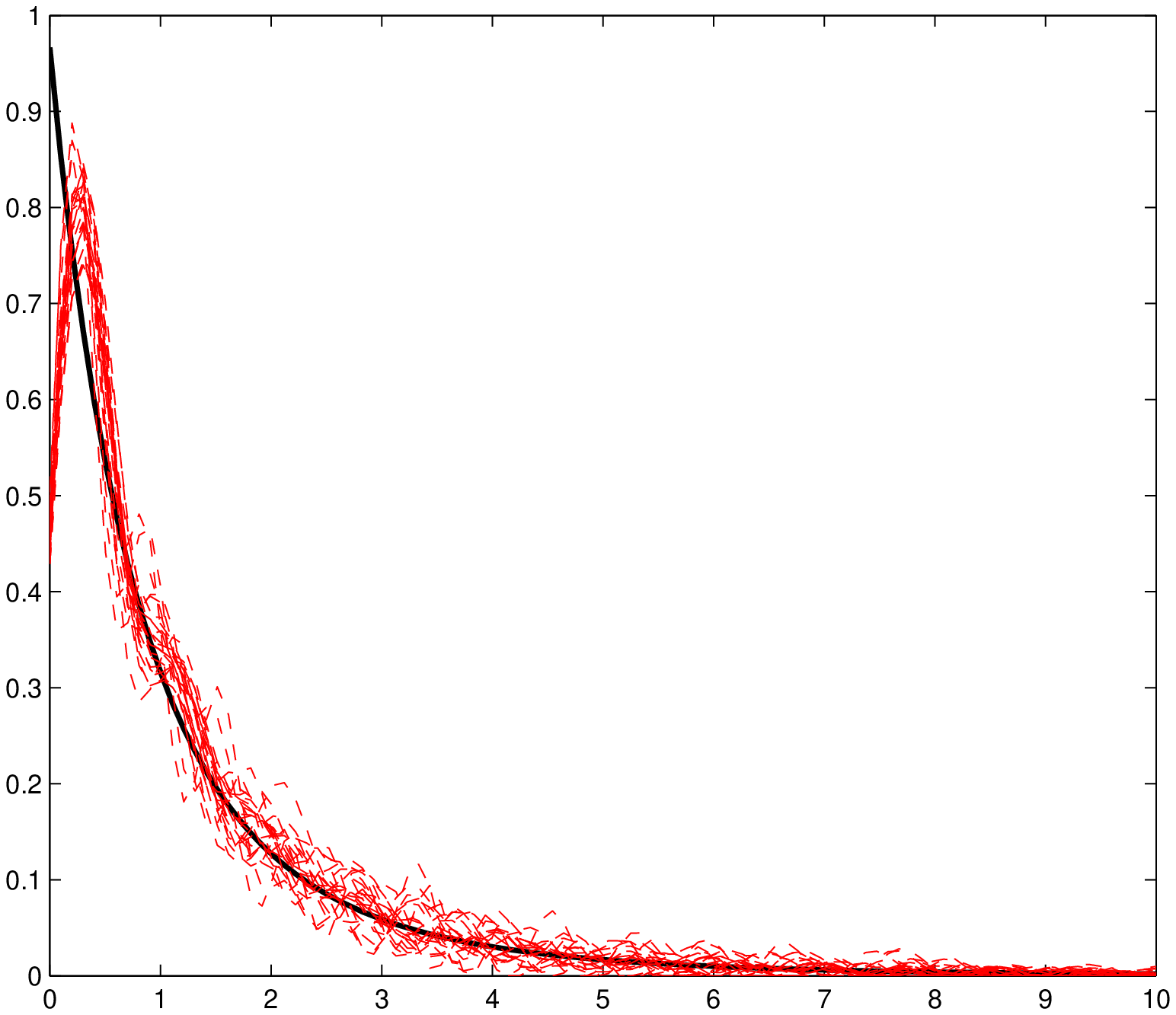}
&\includegraphics[scale=0.15]{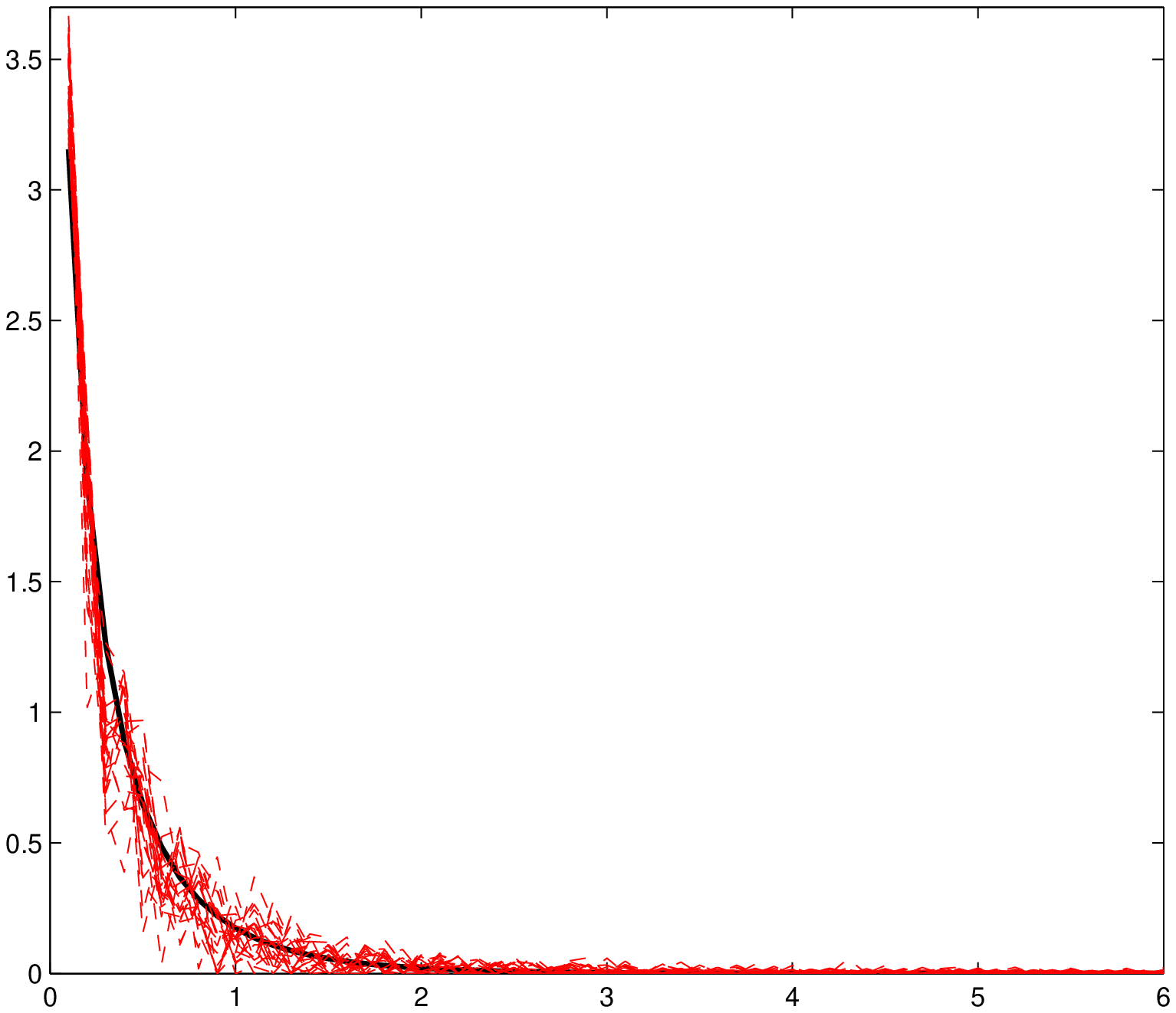}\\
3.00 (0.00)&15.3 (2.0)&33.6 (3.5)&81.0 (7.6)\\
\multicolumn{4}{c}{Estimation with $\sigma=0.1$, $\mu=2$, $n=2000$}\\
\end{tabular}
\caption{True density and 25 estimated curves. Estimation by deconvolution with sinc basis for different noise levels and different levels of the pile-up effect. Numbers below the figures indicate mean and standard deviation of the selected model $\bar m$. }\label{planche_noisy}
\end{figure}
\end{center}

When no noise is added we applied the method described in Section \ref{firststep} with the simple trigonometric basis. The numerical constant $\kappa$ of the penalty (\ref{pen})  is set to 0.5 resulting from a previous calibration by simulation. The Poisson parameter varies from 0.01 over 0.5 to 2. The mean MISE over 25 paths are computed on the intervals of representation.
From Figure \ref{planche_nonoise} one can see that the results are rather good, in spite of small side effects which would be avoided with piecewise polynomial bases. From this point of view all representations in Figure \ref{planche_nonoise} are cut on the right.  We see that the estimator performs  well for a large range of values of the Poisson parameter. The first row corresponds to data where the pile-up effect is negligible, as the Poisson parameter is equal to 0.01, and hence serves as a benchmark. Here estimation errors are mainly due to the choice of a trigonometric basis, that easily recovers the Gamma density while the Weibull density is much harder to approximate in this basis. In the other rows the pile-up effect is considerably increased, however the accuracy is hardly affected and the estimator is still rather stable. The pile-up effect is hence correctly taken into account in the estimation procedure.

\begin{table}
{\footnotesize
\begin{tabular}{ccccccc}
& \multicolumn{6}{c}{Exponential noise} \\\hline\hline
$(\sigma^2, \mu)$& $(0.2, 0.5)$ & $(0.2, 1.5)$ & $(0.2, 2)$ & $(1, 0.5)$ & $(1,1.5)$ & $(1, 2)$ \\
\hline
Gamma& .063 (.042) & .081 (.045) & .112 (.026) & .061 (.039) & .088 (.040) & .115 (.028) \\
    & .063 (.042) & .081 (.045) & .112 (.026) & .061 (.039) & .087 (.040) & .115 (.028) \\
\multicolumn{7}{c}{}\\
Exponential& 1.11 (0.22) & 1.20 (0.26) & 1.45 (0.21) & 1.36 (0.26) & 1.40 (0.24) & 1.67 (0.27) \\
    & 1.11 (0.22) & 1.19 (0.25) & 1.46 (0.21) & 1.36 (0.27) & 1.40 (0.24) & 1.67 (0.27) \\
\multicolumn{7}{c}{}\\
Pareto& 4.25 (0.82) & 4.55 (0.58) & 5.45 (0.84) & 6.62 (1.5) & 6.58 (0.95) & 8.09 (1.2) \\
    & 4.23 (0.83) & 4.56 (0.61) & 5.47 (0.83) & 6.62 (1.6) & 6.58 (1.0) & 8.09 (1.2) \\
\multicolumn{7}{c}{}\\
Weibull& 10.6 (6.7) & 9.46 (5.0) & 9.22 (2.7) & 21.4 (4.1) & 26.7 (5.6) & 39.5 (5.9) \\
    & 8.54 (4.7) & 9.40 (4.8) & 9.30 (2.3) & 22.1 (4.8) & 26.7 (5.7) & 40.1 (5.7) \\
\hline\multicolumn{7}{c}{}\\
& \multicolumn{6}{c}{Bi-exponential noise} \\\hline\hline
$(\sigma^2, \mu)$& $(0.2, 0.5)$ & $(0.2, 1.5)$ & $(0.2, 2)$ & $(1, 0.5)$ & $(1,1.5)$ & $(1, 2)$ \\
\hline
Gamma& .060 (.032) &.075 (.040) &.113 (.023) &.061 (.048) &.088 (.043) &.114 (.025) \\
    & .060 (.032) &.075 (.040) &.113 (.023) &.062 (.048) &.089 (.043) &.114 (.025) \\
\multicolumn{7}{c}{}\\
Exponential& 1.06 (0.20) &1.14 (0.17) &1.49 (0.26) &1.23 (0.27) &1.37 (0.28) &1.62 (0.28) \\
                 & 1.06 (0.20) &1.14 (0.16) &1.48 (0.25) &1.25 (0.26) &1.37 (0.28) &1.62 (0.27) \\
\multicolumn{7}{c}{}\\
Pareto& 4.15 (0.76) &4.31 (0.69) &5.08 (0.71) &6.08 (1.5) &6.41 (1.1) &7.43 (1.0) \\
         & 4.14 (0.77) &4.30 (0.69)  &5.07 (0.72) &6.11 (1.6) &6.49 (1.2) &7.45 (1.1) \\
\multicolumn{7}{c}{}\\
Weibull& 10.2 (6.1) &8.89 (5.6) &8.29 (2.1) &24.7 (3.9) &29.4 (4.5) &40.1 (5.2) \\
           & 8.25 (4.3) &8.75 (5.4) &8.31 (2.2) &24.9 (4.3) &29.5 (4.9) &40.4 (5.3) \\
\hline\multicolumn{7}{c}{}\\
\end{tabular}
\medskip\
\caption{100 $\times$ mean MISE and standard deviation in parentheses.  First lines correspond to exact noise distribution, second lines give results obtained with estimated noise distribution.}\label{tab1}}
\end{table}

The adaptive estimator described in Section \ref{secondstep} is  tested with the numerical constants $\kappa'=1$ and $\kappa''=0.001$ in (\ref{pendec}).  The value of $\kappa''$ is very small and makes the logarithmic term in general negligible except when $c_w^2$ is large (for instance $c_w^2\approx 416$ for $\mu=2$).
 The results are given in Figure \ref{planche_noisy}. Now the observations are $Y=X+\eta$, where $\eta=\sigma \varepsilon$. In the first row, the pile-up effect is almost negligible ($\mu=0.01$), but $\sigma$ is rather large. That is, the first row illustrates the performance of the deconvolution step of the estimation procedure. In contrast, for the last row $\sigma$ is taken to be small, but the pile-up effect is significant ($\mu = 2$), to see how the estimator copes with the pile-up effect. The second row is an intermediate situation, illustrating how the estimator performs when the variance of the noise and the pile-up effect are both non negligible.

The 25 curves indicate variability bands for the estimation procedure. They show that the estimator is quite stable, especially in the last rows. Moreover, the selected model order $\bar m$ is different from one example to the other. Globally the dimension $m$ increases when going from example 1 to 4. That means that the estimator adapts to the peaks that are more and more difficult to recover.

In Table \ref{tab1} the MISE of the estimation procedure is analyzed. The table gives the empirical mean and standard deviation of the MISE obtained over 100 simulated datasets. This is done for the same four examples of distributions as above. We compare the error for the estimator using the exact noise distribution to the estimator based on an approximation of the noise distribution based on an independent noise sample of size 500.
Moreover, we  study the influence of the noise distribution on the estimator.  Therefore, we consider, on the one hand  exponential noise with variances  $\sigma^2\in\{0.2, 1\}$, and on  the other hand density (\ref{dens}) with $\alpha= 2$,  $\beta= 1$, $\nu= 1$, $\tau= 2$ (multiplied with adequate constants to have same variance $\sigma^2$ as for the exponential distributions).

\begin{figure}
\begin{center}\includegraphics[scale=0.2]{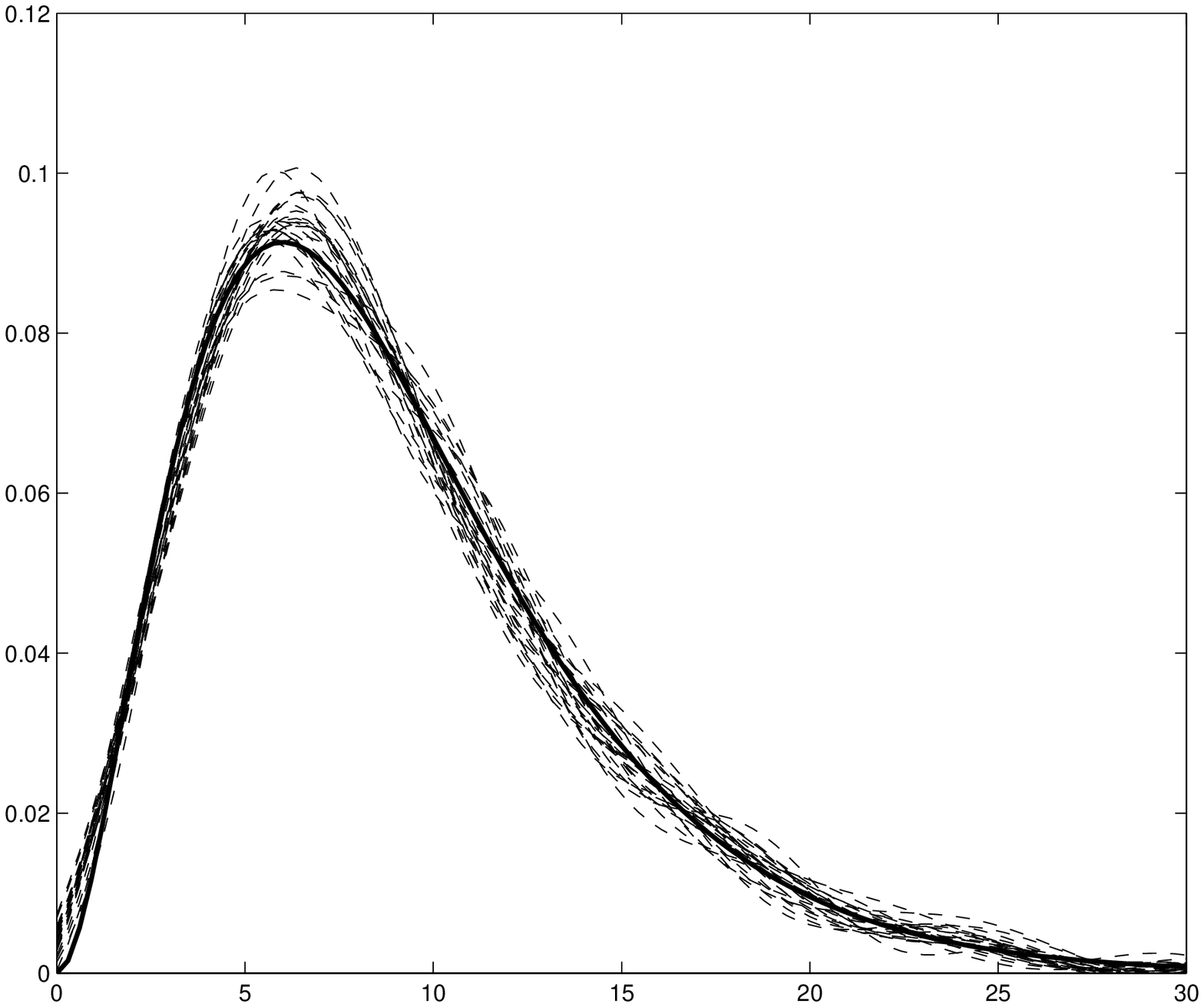}
\includegraphics[scale=0.2]{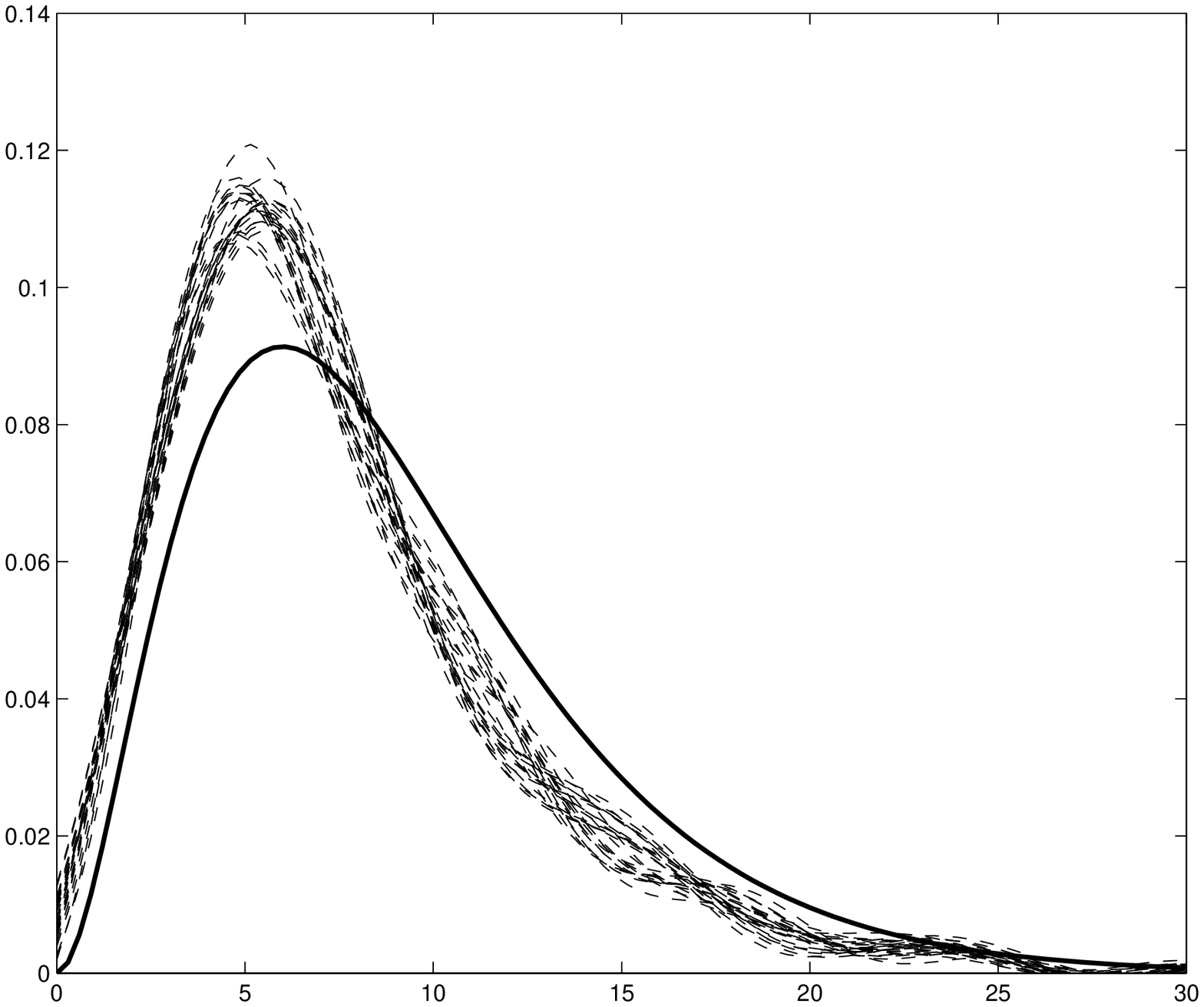}\includegraphics[scale=0.2]{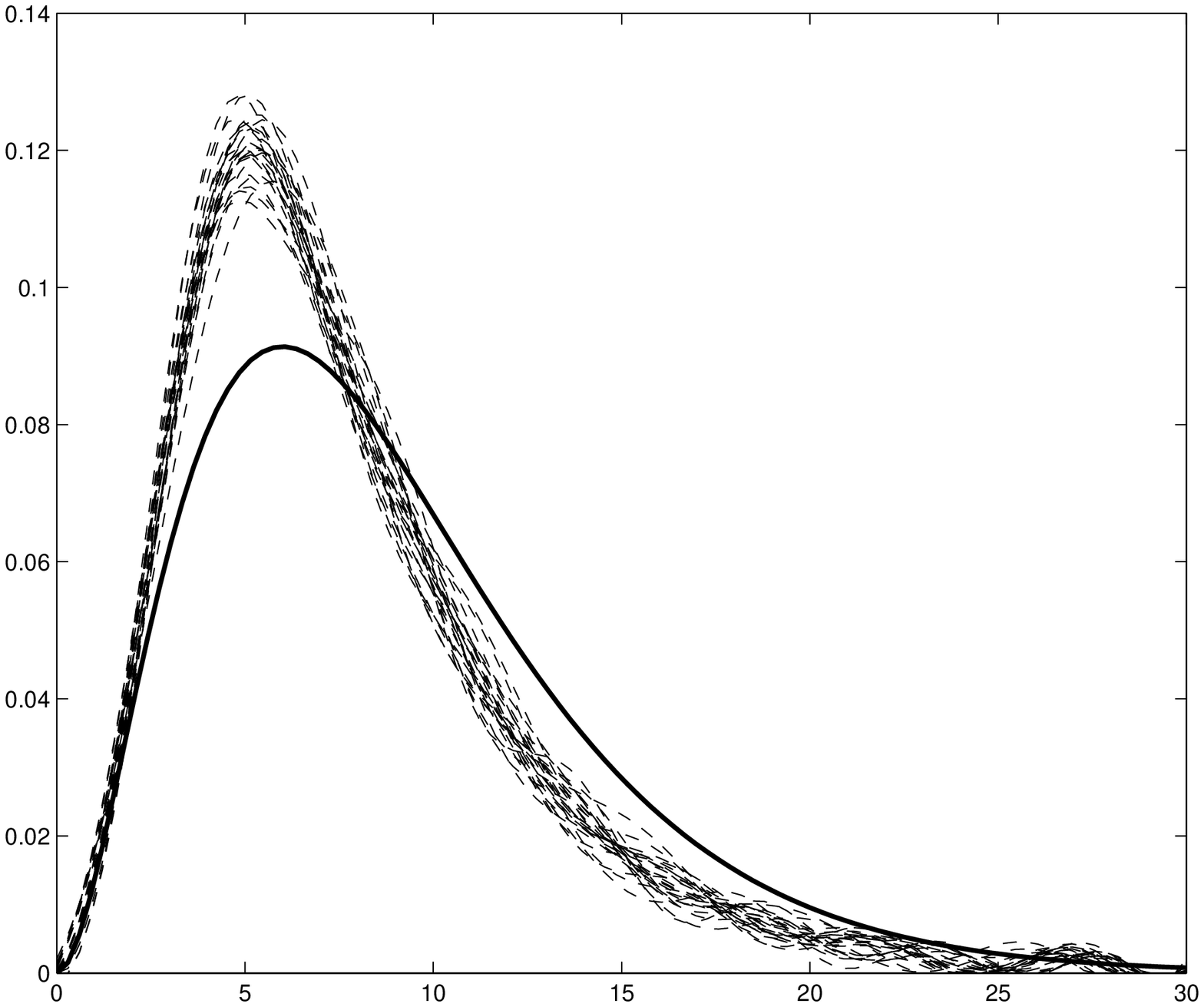}\\
(a) \hspace{4cm} (b) \hspace{4cm} (c)
\caption{(a) Estimation with pile-up correction and deconvolution. (b) No pile-up correction. (c) No deconvolution. }\label{fig_correct}
\end{center}
\end{figure}

From Table \ref{tab1} it is clear that increasing the variance of the noise distribution
increases the error. Furthermore, changing the type of the noise does not influence a lot the estimation procedure.
Indeed, the second case (\ref{dens}) is just slightly less favorable than the exponential distribution. This difference is in accordance with Proposition \ref{rateofdec} that holds with $\gamma=1$ for the exponential and with $\gamma=2$ for the other density. The comparison with the results based on an approximated noise distribution (second lines) reveals that there is rarely a difference between the two methods. Indeed, using an approximation of the noise does not corrupt the results, in some cases we even observe an improvement of the error. We show in Figure \ref{fig_correct} that it is indispensable to take into account both the pile-up correction (which is omitted in (b) where $w(i/n)$ is replaced by $i/n$\,) and the deconvolution correction (which is omitted in (c) where the estimation is done with the method of Section \ref{firststep}  and the trigonometric basis).
Thus, we conclude from these simulation results for the fluorescence setting that it is justified to use an estimate of the noise instead of the theoretical distribution.

\subsection{Application to Fluorescence Measurements}\label{realdata}

\begin{figure}
\begin{center}\includegraphics[scale=0.34]{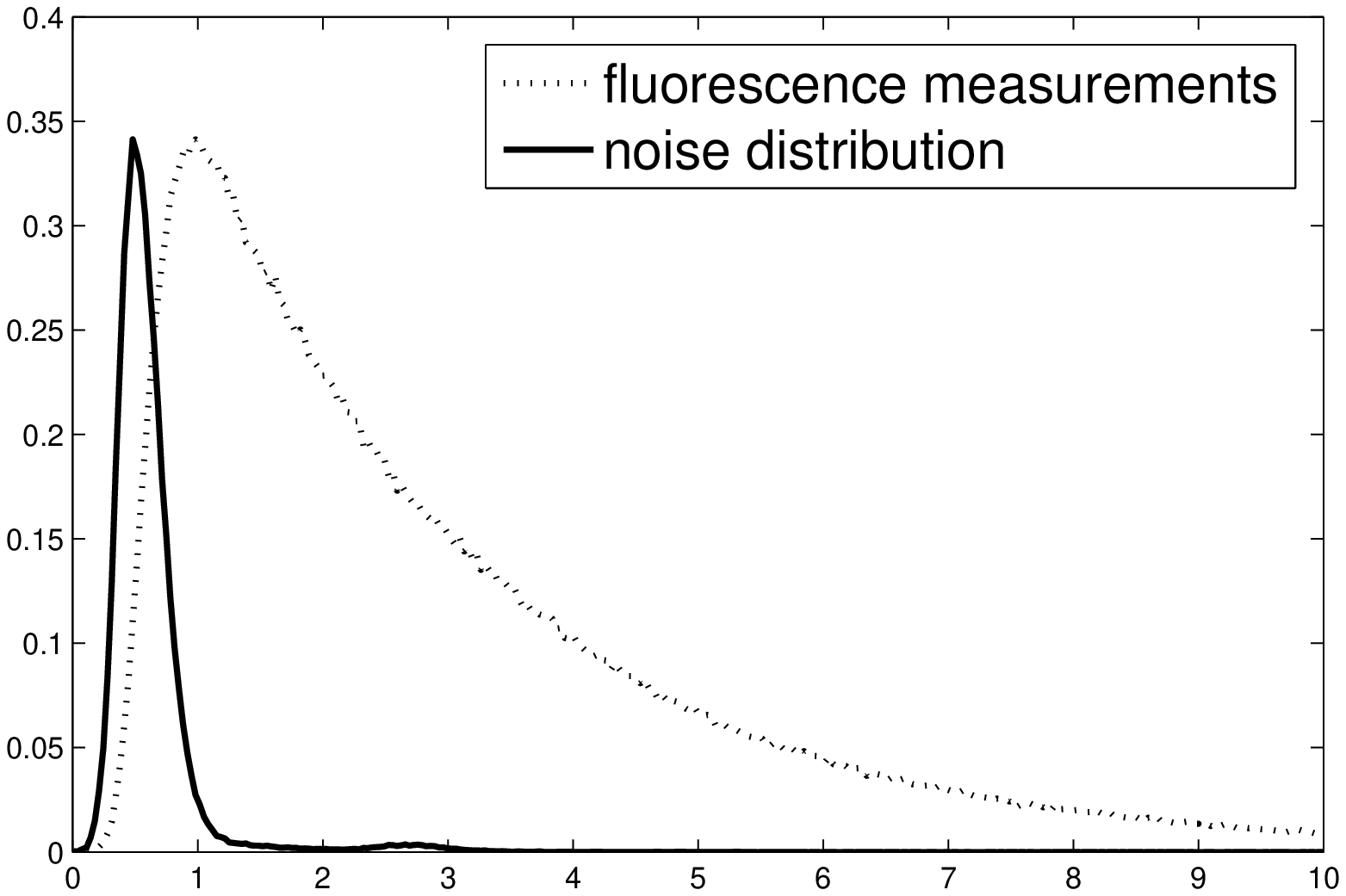}
\includegraphics[scale=0.23]{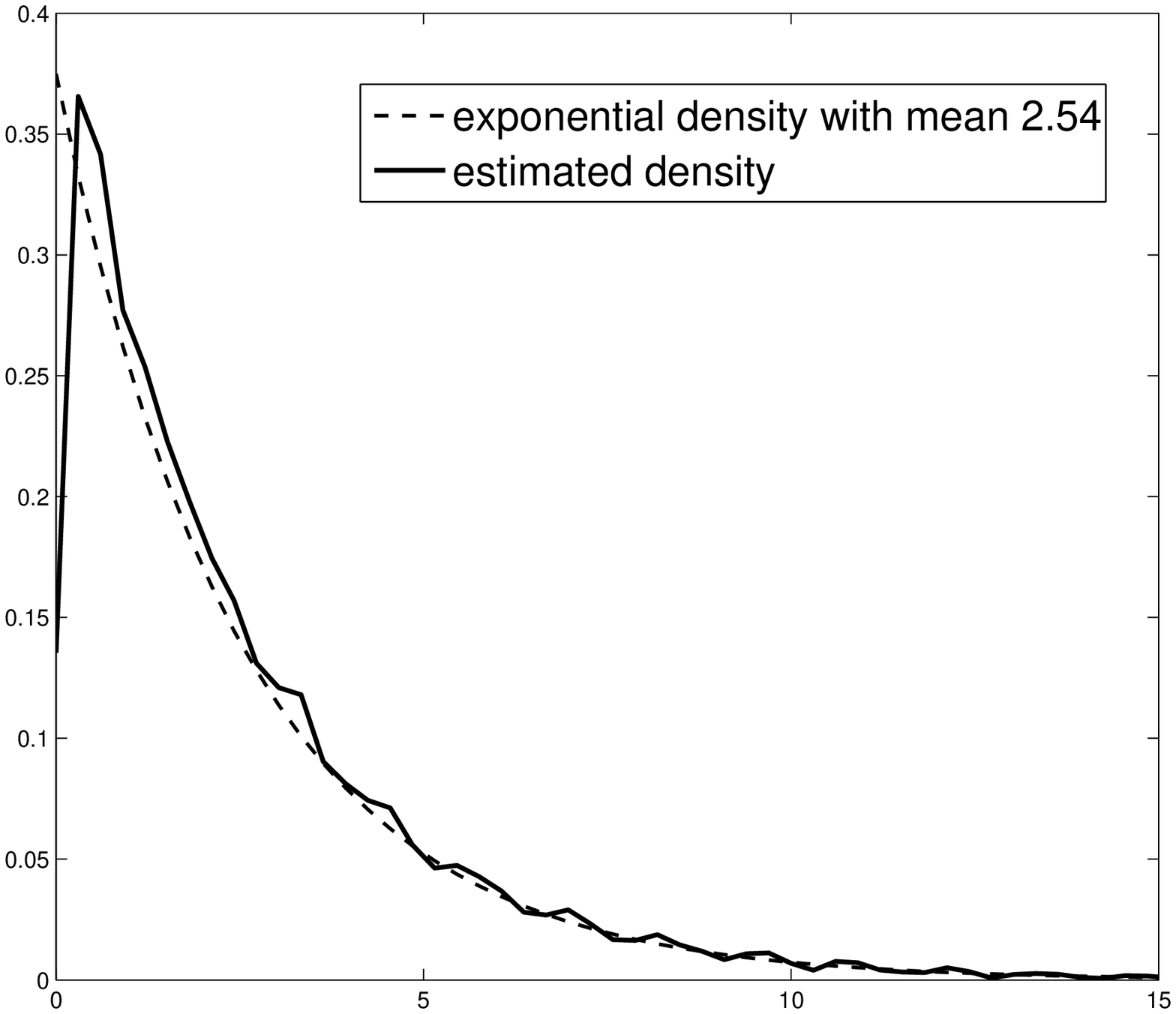}\\
(a) \hspace{5cm} (b)
\caption{(a) Fluorescence lifetime measurements (solid line) and independent sample of the noise distribution (dashed). (b) Density estimator (solid) and `true' exponential density with mean 2.54 (dashed).}\label{fig_picoquant_data}
\end{center}
\end{figure}

We finally applied the estimation procedure to real fluorescence lifetime measurements obtained by TCSPC. The data analyzed here are graphically presented in Figure \ref{fig_picoquant_data} (a) by the histogram of the fluorescence lifetime measurements and the histogram of the noise distribution based on a sample obtained independently from the fluorescence measurements. The sample size of the fluorescence measurements is $n = 1,743,811$. The same sample of the noise distribution has already been considered in Figure~\ref{fitted_density}, where it is compared to the parameterized density given by (\ref{dens}). In this setting the true density is known to be an exponential distribution with mean 2.54 nanoseconds and the Poisson parameter equals 0.166. The knowledge of the true density allows to evaluate the performance of our estimator. More details on the data and their acquisition can be found in \cite{PAT}.

We applied the estimator from Section \ref{secondstep}  with the sinc basis to this dataset. The numerical constants are $\kappa'=1$ and $\kappa''=0.001$. Figure \ref{fig_picoquant_data} (b) shows the estimation result in comparison to the exponential density with mean 2.54. We observe that the estimated function is quite close to the `true' one. This indicates that the estimation procedure takes the errors present in the real data adequately into account and that the modeling by the pile-up distortion and additive measurement errors is appropriate.

We conclude that the estimation methods proposed in this paper have a satisfactory behavior in various settings and give rather good results on both synthetic and real data. Nevertheless, we observed that the performance depends on the choice of the basis and on the smoothness of the target density. Here only two bases are considered, but others should work as well and may improve the results in certain settings.

\section{Proofs}\label{proofs}
\subsection{Proof of Proposition \ref{risksansbruit}.}
Pythagoras formula yields
$\|f-\hat f_m\|^2 = \|f-f_m\|^2 + \|f_m-\hat f_m\|^2.$
 By definition of the orthogonal projection $f_m=\sum_{\lambda\in \Lambda_m}a_\lambda\varphi_\lambda$ and  by using equality (\ref{pileup_property}), we have $a_\lambda=\langle \varphi_\lambda, f_Y\rangle = {\mathbb E}(\varphi_\lambda(Y))={\mathbb E}(\varphi_\lambda(Z_1)w\circ G(Z_1)).$
This, together with formula (\ref{estimbias}) implies that
$\|f_m-\hat f_m\|^2 =\sum_{\lambda \in \Lambda_m} (a_\lambda -\hat a_\lambda)^2.$
If we  define
\begin{equation}\label{nun} \nu_n(h)=\frac 1n \sum_{i=1}^n [h(Z_i)w\circ G(Z_i)- {\mathbb E}(h(Z_i)~w\circ G(Z_i))],
\end{equation}
\begin{equation}\label{rn}
R_n(h)= \frac 1n\sum_{i=1}^nh(Z_i)[w\circ \hat G_n(Z_i)- w\circ G(Z_i)]\;,\end{equation}
then we get
 $\|f_m-\hat f_m\|^2 \leq 2\sum_{\lambda \in \Lambda_m}(\nu_n(\varphi_\lambda)^2+ R_n(\varphi_\lambda)^2).$
We have, on the one hand,
\begin{align}\nonumber
 &\sum_{\lambda\in\Lambda_m} {\mathbb E}(\nu_n^2(\varphi_\lambda))
 =\sum_{\lambda\in\Lambda_m} \frac 1n
{\rm Var}(\varphi_\lambda(Z_i)w\circ G(Z_i)) \leq  \sum_{\lambda\in\Lambda_m} \frac
1n {\mathbb E}\left[\varphi_\lambda^2(Z_1)(w\circ G(Z_1))^2\right]\\
\label{step3}
&\quad\leq   \frac
1n {\mathbb E}\left[ \|\sum_{\lambda\in\Lambda_m}\varphi_\lambda^2\|_{\infty}(w\circ
G(Z_1))^2\right]
\leq  \Phi_0 \frac{D_m}n {\mathbb E}[(w\circ G(Z_1))^2]
\leq \Phi_0w_1^2 \frac{D_m}n\;,
\end{align} because the basis satisfies (\ref{phi0}).
On the other hand, we have
\begin{eqnarray}\nonumber
&& \sum_{\lambda\in\Lambda_m} {\mathbb E}(R_n^2(\varphi_\lambda)) \leq  \sum_{\lambda\in\Lambda_m} {\mathbb E}\left[\left(
\frac 1n\sum_{i=1}^n \varphi_\lambda(Z_i)[w\circ \hat G_n(Z_i)-
w\circ G(Z_i)]\right)^2\right] \\\nonumber  &\leq & \frac 1n\sum_{i=1}^n \sum_{\lambda\in\Lambda_m} {\mathbb E}\left(\varphi_\lambda^2(Z_i)
[w\circ \hat G_n(Z_i)-w\circ G(Z_i)]^2\right) \\\label{step4}
&\leq & c_w^2 \sum_{\lambda\in\Lambda_m} {\mathbb E}\left(\|G-\hat G_n\|_{\infty}^2\varphi_\lambda^2(Z_i)\right)
\leq  c_w^2\Phi_0 D_m {\mathbb E}\left(\|G-\hat G_n\|_{\infty}^2\right) \leq  c_w^2\Phi_0 \frac{D_m}n \end{eqnarray}
with (\ref{lip}) and because of ${\mathbb E}\left(\|G-\hat G_n\|_{\infty}^2 \right) \leq 1/n$ \citep[see e.g.][p. 462]{BC}.
By gathering all terms, we obtain the risk bound stated in Proposition \ref{risksansbruit}. $\Box$

\subsection{Sketch of proof of Theorem \ref{adaptbias}}

We can write
$\gamma_n(t)-\gamma_n(s)=\|t-f_Y\|^2 -\|s-f_Y\|^2 -2\nu_n(t-s)-2R_n(t-s),$
where $\nu_n$ and $R_n$ are defined by (\ref{nun}) and (\ref{rn}).
By definition of $\hat f_{\hat m}$ we have for all $m\in {\mathcal M}_n$, $\gamma_n(\hat f_{\hat m}) + {\rm pen}(\hat m) \leq \gamma_n(f_m) + {\rm pen}(m)$. This can be rewritten as
$\|\hat f_{\hat m}-f_Y\|^2\leq \|f_m-f_Y\|^2+ {\rm pen}(m) + 2\nu_n(\hat f_{\hat m}-f_m)- {\rm pen}(\hat m) +2 R_n(\hat f_{\hat m}-f_m) .$
Using this and
and that $2xy\leq x^2/\theta + \theta y^2$ for all nonnegative $x,y, \theta$, we obtain
\begin{align*}
\|f_Y-\hat f_{\hat m}\|^2
&\leq  \|f_Y-f_m\|^2 +{\rm pen}(m)+ 2\nu_n(\hat f_{\hat m} - f_m)- {\rm pen}(\hat m) + 2R_n(\hat f_{\hat m} -f_m)
\end{align*}
\begin{align*}
\|f_Y-\hat f_{\hat m}\|^2 &\leq   \|f_Y-f_m\|^2+{\rm pen}(m) + 2\|\hat f_{\hat m } - f_m\| \sup_{t\in S_{\hat m}+S_m, \|t\|=1}
|\nu_n(t)| -{\rm pen}(\hat m)\\
&+ 2\|\hat f_{\hat m} -f_m\|\sup_{t\in S_{\hat m}+S_m,\|t\|=1}|R_n(t)|\\
&\leq   \|f_Y-f_m\|^2 +{\rm pen}(m)+  \frac 14\|\hat f_{\hat m} - f_m\|^2 + 2 \sup_{t\in S_{\hat m}+S_m, \|t\|=1}
[\nu_n(t)]^2\\
&- {\rm pen}(\hat m)+ \frac 18 \|\hat f_{\hat m} -f_m\|^2 + 8\sup_{t\in S_{\hat m}+ S_m,\|t\|=1}[R_n(t)]^2\;.
\end{align*}

As $\|\hat f_{\hat m}-f_m\|^2 \leq 2( \|\hat f_{\hat m}-f\|^2 + \|f_m-f\|^2)$, this yields
\begin{eqnarray*} \frac 14 {\mathbb E}[ \|f-\hat f_{\hat m}\|^2]&\leq &\frac 74 \|f-f_m\|^2 + 2{\rm pen}(m)+  8{\mathbb E}\left( \sup_{t\in S_{m_n},\|t\|=1}[R_n(t)]^2\right)
\\ &&
+4{\mathbb E}\left( \sup_{t\in S_{\hat m}+S_m, \|t\|=1} [\nu_n(t)]^2- ({\rm pen}(m)+{\rm pen}(\hat m))/4\right)_+\;.
\end{eqnarray*}

Then  the term ${\mathbb E}\left( \sup_{t\in S_{\hat m}+S_m, \|t\|=1}
[\nu_n(t)]^2- ({\rm pen}(m)+{\rm pen}(\hat m))/4\right)_+$ is bounded by $C/n$ by using Talagrand Inequality in a standard way \cite[see e.g.][]{BCG}.
For the last term ${\mathbb E}\left( \sup_{t\in
S_{m_n},\|t\|=1}[R_n(t)]^2\right)$, we define $\Omega_G$ by  \begin{equation}\label{omegag}\Omega_G={\{ \sqrt{n} \|\hat G_n-G\|_{\infty}\leq \sqrt{\ln(n)}\}}.\end{equation}  Now, we know from \cite{MAS1} that
\begin{equation}\label{massart} {\mathbb P}(\sqrt{n}\|\hat G_n-G\|_{\infty}\geq \lambda)\leq 2e^{-2\lambda^2}.\end{equation}
This implies that ${\mathbb P}(\Omega_G^c)\leq 2/n^2$.
Then we write that ${\mathbb E}\left( \sup_{t\in S_{m_n},\|t\|=1}[R_n(t)]^2\right)$ is less than 
\begin{eqnarray*}{\mathbb E}\left( \sup_{t\in
S_{m_n},\|t\|=1}[R_n(t)\1_{\Omega_G}]^2\right)+ {\mathbb E}\left( \sup_{t\in
S_{m_n},\|t\|=1}[R_n(t)\1_{\Omega_G^c}]^2\right) && :={\mathcal R}_1+{\mathcal R}_2.\end{eqnarray*}
For the first term, we have
\begin{eqnarray*} {\mathcal R}_1 & \leq &  c_w^2 {\mathbb E}\left[\|\hat G_n-G\|^2_\infty \1_{\Omega_G}{\mathbb E}\left( \sup_{t\in S_{m_n},\|t\|=1}( \frac 1n\sum_{i=1}^n |t(Z_i)|)^2\right)\right]
 \\ &\leq & c_w^2 \frac{\ln(n)}n {\mathbb E}\left( \sup_{t\in S_{m_n},\|t\|=1}(\frac 1n\sum_{i=1}^n t^2(Z_i))\right)\\
\\ &\leq & 2c_w^2 \frac{\ln(n)}n \left[{\mathbb E}\left( \sup_{t\in S_{m_n},\|t\|=1}|\nu'_n(t^2)|\right)+
\sup_{t\in S_{m_n},\|t\|=1}{\mathbb E}(t^2(Z_1))\right]
\end{eqnarray*}
where $\nu'_n(t)=\frac 1n\sum_{i=1}^n (t(Z_i)-{\mathbb E}(t(Z_1))$. It is proved in \cite{BC} that
${\mathbb E}\left( \sup_{t\in S_{m_n},\|t\|=1}|\nu'_n(t^2)|\right)\leq  C\ln(n)$ if the density of $Z_1$ is bounded and $N_n\leq O(n)$ for bases [DP] and [W] and $N_n\leq O(\sqrt{n})$ for basis [T]. Moreover ${\mathbb E}(t^2(Z_1))\leq \|t\|^2\|f_Y\|_\infty/w_0$. We obtain ${\mathcal R}_1\leq C\ln^2(n)/n$.
On the other hand, we have
\begin{eqnarray*}
{\mathcal R}_2 &\leq & \sum_\lambda {\mathbb E}(R_n^2(\varphi_\lambda)\1_{\Omega^c})\leq
c_w^2\Phi_0n {\mathbb E}^{1/2}(\|\hat G_n-G\|^4_\infty){\mathbb P}^{1/2}(\Omega_G^c) \leq \frac Cn.\end{eqnarray*}
This yields ${\mathbb E}\left( \sup_{t\in
S_{m_n},\|t\|=1}[R_n(t)]^2\right)\leq C\ln^2(n)/n$.
Finally we obtain that, for all $m\in {\mathcal M}_n$,
${\mathbb E}[ \|f-\hat f_{\hat m}\|^2]\leq  7 \|f-f_m\|^2 + 8{\rm pen}(m)+ K{\ln^2(n)}/n$, which ends the proof.  $\Box$

\subsection{Proof of Proposition \ref{borne}.}
We have $\|\bar f_m-f_Y\|^2={(2\pi)^{-1}} \|\bar f_m^*-f_Y^*\|^2 = (2\pi)^{-1}(\| \bar f_m^*-f_{Y,m}^*\|^2 + \|f_{Y,m}^*-f_Y^*\|^2).$
\begin{align}\nonumber
&\| \bar f_m^*-f_{Y,m}^*\|^2
= \int_{-\pi m}^{\pi m} \frac{\rmd u}{|f_\eta^*(u)|^2} \frac 1{n^2}\left| \sum_{k=1}^n \left[e^{-iuZ_k} w\circ \hat G_n(Z_k) - {\mathbb E}(e^{-iuZ_k} w\circ  G(Z_k)) \right]\right|^2 \\
\nonumber
&\quad \leq  2 \int_{-\pi m}^{\pi m} \frac{\rmd u}{|f_\eta^*(u)|^2} \frac 1{n^2}\left| \sum_{k=1}^n \left[e^{-iuZ_k} w\circ \hat G_n(Z_k)-e^{-iuZ_k} w\circ G(Z_k)\right]\right|^2 \\
\label{toto}
& \qquad+  2 \int_{-\pi m}^{\pi m} \frac{\rmd u}{|f_\eta^*(u)|^2} \frac 1{n^2}\left| \sum_{k=1}^n \left[e^{-iuZ_k} w\circ G(Z_k)-{\mathbb E}(e^{-iuZ_k} w\circ G(Z_k)) \right]\right|^2\;.
\end{align}
The expectation of the first term on the right-hand side of (\ref{toto}) is less than or equal to
\begin{eqnarray*} &&\frac 2n  \sum_{k=1}^n  \int_{-\pi m}^{\pi m} \frac{\rmd u}{|f_\eta^*(u)|^2}{\mathbb E}(| w\circ \hat G_n(Z_k)-w\circ G(Z_k)|^2)\\
&\leq & c_w^2 {\mathbb E}\left(\|\hat G_n-G\|^2_{\infty}\right)\int_{-\pi m}^{\pi m}\rmd u / |f_\eta^*(u)|^2
\leq 2\pi c_1c_w^2\frac{\Delta_\eta(m)}n\;,\end{eqnarray*}
by using ${\mathbb E}(\|\hat G_n -G\|^{2k}_{\infty})\leq c_k/n^k$ (see e.g. Lemma 6.1 p.
462, \cite{BC}   which is a straightforward consequence of \cite{MAS1}). Here $c_k$ is a numerical constant that depends on $k$ only.
The expectation of the second term on the right-hand side of (\ref{toto}) is a variance and less than or equal to
$$\frac 2n \int_{-\pi m}^{\pi m} \frac{\rmd u}{|f_\eta^*(u)|^2} {\rm Var}(e^{-iuZ_1} w\circ G(Z_1))
 \leq 4\pi \frac{\Delta_\eta(m) {\mathbb E}[(w\circ G(Z_1))^2]}n\;.$$
Gathering the terms completes the proof of Proposition \ref{borne}. $\Box$

\subsection{Proof of Theorem \ref{adaptconv}.}

We have the following decomposition of the contrast for functions  $s,t$ in $\bar S_m$,
\begin{equation}\label{decgam}
\gamma_n^\dagger (t)-\gamma_n^\dagger (s)= \|t-f_Y\|^2 -\|s-f_Y\|^2-2\bar \nu_n(t-s) -2 \bar R_n(t-s)\;,
\end{equation}
where
\begin{equation}\label{nun2}
\bar \nu_n(t) =\frac 1{2\pi n} \sum_{k=1}^n \int \frac{t^*(-u)\left[ e^{-iuZ_k}(w\circ G)(Z_k) - {\mathbb
E}(e^{-iuZ_k}(w\circ G)(Z_k))\right]}{f_\eta^*(u)} \rmd u \;,
\end{equation}
and
\begin{equation}\label{rn2} \bar R_n(t) =\frac 1{2\pi n} \sum_{k=1}^n \int\frac{t^*(-u) e^{-iuZ_k}}{f_\eta^*(u)} \rmd u \; [(w\circ \hat G_n)(Z_k) -
(w\circ G)(Z_k)]\;.
\end{equation}

We start with decomposition (\ref{decgam}). We take $t=\bar f_{\bar m}$ and
$s=f_{Y,m}$.  Since
$\gamma_n^\dagger (\bar f_{\bar m})+ \overline{{\rm pen}}(\bar m)  \leq  \gamma_n^\dagger(f_m)+
\overline{{\rm pen}}(m)$, we get
\begin{eqnarray}\nonumber  \frac 14{\mathbb E}[ \|f_Y-\bar f_{\bar m}\|^2]
&\leq & \frac 74 \|f_Y-f_{Y,m}\|^2 + \overline{{\rm pen}}(m) + 4{\mathbb E}\left(
\sup_{t\in B_{m,\bar m}} [\bar \nu_n(t)]^2\right) - {\mathbb E}(\overline{{\rm pen}}(\bar m)) \\ \label{etap3}
&& \hspace{1cm}  + 8{\mathbb E}\left( \sup_{t\in B_{m,\bar m}}
[\bar R_n(t)]^2 \right), \end{eqnarray} where $\bar\nu_n(t)$ and $\bar R_n(t)$ are
defined by (\ref{nun2}) and (\ref{rn2}) and
$B_{m}=\{t \in \bar S_m, \|t\|=1\}, $ and $B_{m, m'}=\{t \in \bar S_m +\bar S_{m'}, \|t\|=1\}.$
Following a classical application of Talagrand Inequality in the deconvolution context for ordinary smooth noise \citep{CRT}, we
deduce the following Lemma.
\begin{lem}\label{controlnun}
Under the Assumptions of Theorem \ref{adaptconv},
$${\mathbb E}\left( \sup_{t\in B_{m,\bar m}} [\bar \nu_n(t)]^2 - p_1(m, \bar m) \right)_+ \leq \frac cn,$$
where $p_1(m,  m')=2{\mathbb E}((w\circ G)^2(Z_1))\Delta_\eta(m\vee m')/n=2(\int_0^1w^2(u)\rmd u)\Delta_\eta(m\vee m')/n.$
\end{lem}
Moreover for the study $\bar R_n(t)$ we have the following Lemma.
\begin{lem}\label{controlrn}
Under the assumptions of Theorem \ref{adaptconv},
$${\mathbb E}\left( \sup_{t\in B_{m,\bar m}} [\bar R_n(t)]^2 - p_2(m, \bar m) \right) \leq 0,$$
where $p_2(m,m')= c_w^2\Delta_\eta(m\vee m')\ln(n)/n$.
\end{lem}
It follows from the definition of $p_i(m, m')$, $i=1,2$, that there exist  numerical constants $\kappa'$ and $\kappa''$, namely $\kappa', \kappa''\geq 8$, such that $4p_1(m,m')+8p_2(m,m')\leq \overline{{\rm pen}}(m) + \overline{{\rm pen}}(m')$.

Now, starting from (\ref{etap3}), we get, by applying Lemmas \ref{controlnun} and \ref{controlrn},
\begin{align*}
&\frac 14{\mathbb E}[ \|f_Y-\bar f_{\bar m}\|^2]
\leq  \frac 74 \|f_Y-f_{Y,m}\|^2 + \overline{{\rm pen}}(m) + 4{\mathbb E}\left(
\sup_{t\in B_{m,\bar m}} [\bar \nu_n(t)]^2-p_1(m,\bar m)\right)_+
 \\
&\qquad + 8{\mathbb E}\left( \sup_{t\in B_{m,\bar m}} [\bar R_n(t)]^2-p_2(m,\bar m) \right)  + {\mathbb E}[4p_1(m, \bar m)+8 p_2(m,\bar m) - \overline{{\rm pen}}(\bar m)] \\
&\quad\leq  \frac 74 \|f_Y-f_{Y,m}\|^2 + 2\overline{{\rm pen}}(m)  + \frac{c}n.\end{align*}
Therefore if  $\kappa\geq 16$, we get
$ (1/4){\mathbb E}[ \|f_Y-\bar f_{\bar m}\|^2]
\leq  (7/4) \|f_Y-f_{Y,m}\|^2 + 2\overline{{\rm pen}}(m) + c/n.$ This completes the proof of Theorem \ref{adaptconv}.  $\Box$\\

\noindent {\it Proof of Lemma \ref{controlrn}.}
First we remark that, with Cauchy-Schwarz inequality, we have
\begin{eqnarray*}
 |\bar R_n(t)|^2 &\leq &  \frac 1{4\pi^2}\left| \int \frac{t^*(-u)}{f_\eta^*(u)} \left(\frac 1n \sum_{k=1}^n e^{-iuZ_k}[(w\circ \hat G_n)(Z_k) -
(w\circ G)(Z_k)]\rmd u\right) \right|^2 \\ &\leq &  \frac 1{4\pi^2} \int |t^*(u)|^2\rmd u \int_{-\pi m\vee \bar m}^{\pi m\vee \bar m}\frac{\rmd u}{|f_\eta^*(u)|^2}\left(\frac 1n \sum_{k=1}^n
|(w\circ \hat G_n)(Z_k) - (w\circ G)(Z_k)|^2\right).\end{eqnarray*}

Then Parseval Formula gives $\|t^*\|^2=2\pi \|t\|^2$ and we find
$$\sup_{t\in B_{m,\bar m}} |\bar R_n(t)|^2\leq c_w^2 \Delta_\eta(m\vee \bar m)\left(\frac 1n \sum_{k=1}^n
|\hat G_n(Z_k) -  G(Z_k)|^2\right) \leq c_w^2 \Delta_\eta(m\vee \bar m)\|\hat G_n-\hat G\|_\infty^2.$$
Now, we write  $\sup_{t\in B_{m,\bar m}} |\bar R_n(t)|^2= {\mathcal R}_1+{\mathcal R}_2$ by inserting again the indicator functions $\1_{\Omega_G}$ and $\1_{\Omega_G^c}$ where $\Omega_G$ is defined by (\ref{omegag}).
Therefore
\begin{align} \nonumber
&{\mathbb E}\left( \sup_{t\in B_{m,\bar m}} [\bar R_n(t)]^2 -p_2(m,\bar m)\right) \leq
{\mathbb E}({\mathcal R}_1-p_2(m,\bar m)) + {\mathbb E}({\mathcal R}_2)\\ \label{decoup}
&\qquad\leq c_w^2 {\mathbb E}\left(\Delta_\eta(m\vee \bar m)(\|\hat G_n-\hat G\|_\infty^2\1_{\Omega_G}-\frac{\ln(n)}n)\right) \\ \nonumber
&\qquad\quad + cw^2\Delta_\eta(m_n){\mathbb E}(\|\hat G_n-G\|_\infty^2\1_{\Omega_G^c}).
\end{align}
Next $(\|\hat G_n-\hat G\|_\infty^2\1_{\Omega_G}-\ln(n)/n))\leq 0$ by definition of $\Omega_G$ for the first right-hand-side term of (\ref{decoup}). For the second term,
$\Delta(m_n)\leq n$ by the  definition of $m_n$, $\|\hat G_n-G\|_\infty\leq 1$ and it follows from (\ref{massart}) that  ${\mathbb P}(\Omega_G^c)\leq 2/n^2$. Therefore
$${\mathbb E}\left( \sup_{t\in B_{m,\bar m}} [\bar R_n(t)]^2 -p_2(m,\bar m)\right) \leq c_w^2 n {\mathbb P}(\Omega_G^c)\leq 2c_w^2/n.$$

Gathering the bounds gives the result of Lemma \ref{controlrn}. $\Box$

\bibliographystyle{jasa}
\bibliography{biblio}

\begin{thebibliography}{}
\newcommand{\enquote}[1]{``#1''}

\bibitem[Barron et~al.(1999)Barron, Birg\'e, and Massart]{BBM}
Barron, A., Birg\'e, L., and Massart, P. (1999), \enquote{Risk bounds for model
  selection via penalization,} \emph{Probability Theory and Related Fields},
  113, 301--413.

\bibitem[Berberan-Santos et~al.(2005a)Berberan-Santos, Bodunov, and
  Valeur]{BBVa}
Berberan-Santos, M.~N., Bodunov, E.~N., and Valeur, B. (2005a),
  \enquote{Mathematical functions for the analysis of luminescence decays with
  underlying distributions 1. {K}ohlrausch decay function (stretched
  exponential),} \emph{Chemical Physics}, 515, 171--182.

\bibitem[Berberan-Santos et~al.(2005b)Berberan-Santos, Bodunov, and
  Valeur]{BBVb}
Berberan-Santos, M.~N., Bodunov, E.~N., and Valeur, B. (2005b),
  \enquote{Mathematical functions for the analysis of luminescence decays with
  underlying distributions: 2. {B}ecquerel (compressed hyperbola) and related
  decay functions,} \emph{Chemical Physics}, 317, 57--62.

\bibitem[Birg{\'e} and Massart(1997)Birg{\'e} and Massart]{BM}
Birg{\'e}, L. and Massart, P. (1997), \enquote{From model selection to adaptive
  estimation,} in \emph{Festschrift for {L}ucien {L}e {C}am}, pp. 55--87,
  Springer, New York.

\bibitem[Brunel and Comte(2005)Brunel and Comte]{BC}
Brunel, E. and Comte, F. (2005), \enquote{Penalized contrast estimation of
  density and hazard rate with censored data,} \emph{Sankhya}, 67, 441--475.

\bibitem[Brunel et~al.(2005)Brunel, Comte, and Guilloux]{BCG}
Brunel, E., Comte, F., and Guilloux, A. (2005), \enquote{Nonparametric density
  estimation in presence of bias and censoring,} \emph{Test}, 18, 166--194.

\bibitem[Comte and Lacour(2009)Comte and Lacour]{CL}
Comte, F. and Lacour, C. (2009), \enquote{Data driven density estimation in
  presence of unknown convolution operator,} preprint available at
  {http://www.math-info.univ-paris5.fr/map5/Prepublications-2008}.

\bibitem[Comte et~al.(2006)Comte, Rozenholc, and Taupin]{CRT}
Comte, F., Rozenholc, Y., and Taupin, M.-L. (2006), \enquote{Penalized contrast
  estimator for adaptive density deconvolution,} \emph{Canadian Journal of
  Statistics}, 34, 431--452.

\bibitem[Daubechies(1992)Daubechies]{DAU}
Daubechies, I. (1992), \enquote{Ten lectures on wavelets,} in \emph{CBMS-NSF
  Regional Conference Series in Applied Mathematics}, Philadelphia, PA, Society
  for Industrial and Applied Mathematics (SIAM).

\bibitem[Donoho et~al.(1996)Donoho, Johnstone, Kerkyacharian, and Picard]{DJKP}
Donoho, D.~L., Johnstone, I.~M., Kerkyacharian, G., and Picard, D. (1996),
  \enquote{Density estimation by wavelet thresholding,} \emph{Annals of
  Statistics}, 24, 508--539.

\bibitem[Lakowicz(1999)Lakowicz]{LAKO}
Lakowicz, J.~R. (1999), \emph{Principles of {F}luorescence {S}pectroscopy},
  Academic/Plenum, New York.

\bibitem[Massart(1990)Massart]{MAS1}
Massart, P. (1990), \enquote{The tight constant in the
  Dvoretzky-Kiefer-Wolfowitz inequality,} \emph{Annals of Probability}, 18,
  1269--1283.

\bibitem[Meyer(1990)Meyer]{MEY}
Meyer, Y. (1990), \emph{Ondelettes et op\'erateurs}, Hermann, Paris.

\bibitem[O'Connor and Phillips(1984)O'Connor and Phillips]{OCP}
O'Connor, D.~V. and Phillips, D. (1984), \emph{Time-correlated single photon
  counting}, Academic Press, London.

\bibitem[Patting et~al.(2007)Patting, Wahl, Kapusta, and Erdmann]{PAT}
Patting, M., Wahl, M., Kapusta, P., and Erdmann, R. (2007), \enquote{Dead-time
  effects in {TCSPC} data analysis,} in \emph{Proceedings of SPIE}, vol. 6583.

\bibitem[Rebafka et~al.(2010)Rebafka, Roueff, and Souloumiac]{RRS1}
Rebafka, T., Roueff, F., and Souloumiac, A. (2010), \enquote{A corrected
  likelihood approach for the pile-up model with application to fluorescence
  lifetime measurements using exponential mixtures,} \emph{The International
  Journal of Biostatistics}, 6.

\bibitem[Rebafka et~al.(2011)Rebafka, Roueff, and Souloumiac]{RRS2}
Rebafka, T., Roueff, F., and Souloumiac, A. (2011), \enquote{Information bounds
  and {MCMC} parameter estimation for the pile-up model,} \emph{Journal of
  Statistical Planning and Inference}, 141, 1--16.

\bibitem[Tsodikov(2003)Tsodikov]{tsodikov03}
Tsodikov, A. (2003), \enquote{A Generalized Self-Consistency Approach,}
  \emph{Journal of the Royal Statistical Society. Series B (Statistical
  Methodology)}, 65, 759--774.

\bibitem[Valeur(2002)Valeur]{VAL}
Valeur, B. (2002), \emph{Molecular {F}luorescence}, Wiley-VCH, Weinheim.

\end{thebibliography}

\end{document}